\documentclass[graybox, nosecnum]{svmult}

\usepackage{mathptmx}       
\usepackage{helvet}         
\usepackage{courier}        
\usepackage{type1cm}        
\usepackage{makeidx}         
\usepackage{graphicx}        
\usepackage{multicol}        
\usepackage[bottom]{footmisc}
\usepackage{hyperref}        
\usepackage{soul}            
\hypersetup{colorlinks=true,urlcolor=blue}
\usepackage{amsmath}            
\usepackage{amssymb}            

\usepackage[square,numbers]{natbib}

\def\apj{Astrophys. J.}
\def\apjl{Astrophys. J. Lett.}

\def\aap{Astron. Astrophys. }

\def\mnras{Mon. Not. Roy. Astron. Soc. }

\def\prl{Phys. Rev. Lett.}
\def\prd{Phys. Rev. D.}

\def\apss{Astrophys. Space Sci.}

\def\eg{\textit{e.g.}, }

\def\dd{\mathrm{d}}

\makeindex             

\begin{document}
\title*{Gravitational Waves from Core-Collapse Supernovae}
\author{Ernazar Abdikamalov, Giulia Pagliaroli, and David Radice}
\institute{Ernazar Abdikamalov \at %
Department of Physics, School of Sciences and Humanities, Nazarbayev University, Nur-Sultan 010000, Kazakhstan\\
Energetic Cosmos Laboratory, Nazarbayev University, Nur-Sultan 010000, Kazakhstan
\email{ernazar.abdikamalov@nu.edu.kz}
\and Giulia Pagliaroli \at %
Gran Sasso Science Institute, 67100 L’Aquila, Italy, \email{giulia.pagliaroli@gssi.it}
\and David Radice \at %
Institute for Gravitation and the Cosmos, The Pennsylvania State University, University Park, PA 16802, USA, \email{david.radice@psu.edu} \\
Department of Physics, The Pennsylvania State University, University Park, PA 16802, USA \\
Department of Astronomy \& Astrophysics, The Pennsylvania State University, University Park, PA 16802, USA
}
%
%
\maketitle
\abstract{
We summarize our current understanding of gravitational wave emission from core-collapse supernovae. We review the established results from multi-dimensional simulations and, wherever possible, provide back-of-the-envelope calculations to highlight the underlying physical principles. The gravitational waves are predominantly emitted by protoneutron star oscillations. In slowly rotating cases, which represent the most common type of the supernovae, the oscillations are excited by multi-dimensional hydrodynamic instabilities, while in rare rapidly rotating cases, the protoneutron star is born with an oblate deformation due to the centrifugal force. The gravitational wave signal may be marginally visible with current detectors for a source within our galaxy, while future third-generation instruments will enable more robust and detailed observations. The rapidly rotating models that develop non-axisymmetric instabilities may be visible up to a megaparsec distance with the third-generation detectors. Finally, we discuss strategies for multi-messenger observations of supernovae.
}

\section{Keywords} 
Gravitational waves, Core-collapse Supernovae, 
Convection, Shock, SASI, Protoneutron star, Oscillations, Multi-messenger, Neutrinos.

\section{Introduction}
\label{sec:intro}

Core-collapse supernovae (CCSNe), the spectacular explosions of massive stars, play an important role in the evolution of the universe. As the explosive burst sweeps through the interstellar medium, it disseminates newly synthesized elements, profoundly influencing star formation and evolution. They give birth to neutron stars and, in some cases, black holes (BHs). Despite their importance, and after decades of significant research progress, there are still many things that are not understood about CCSNe.

In CCSNe, all the four fundamental forces of nature cooperate in a particular way to produce an explosion. The collapse of the core to a protoneutron star (PNS) releases ${\sim} 10^{53}\,\mathrm{erg}$ gravitational binding energy. Most of it ($\sim 99\%$) escape as neutrinos, while $\sim 1\%$ goes into the kinetic energy of the explosion. Within  $\lesssim 1 \,\mathrm{s}$ after core collapse, powerful aspherical flows develop in the central few hundred kilometers, a region often referred to as the supernova central engine. Besides playing a crucial role in powering the explosion, these flows generate strong gravitational waves (GWs) with energies up to ${\sim} 10^{46}{-}10^{47}\,\mathrm{erg}$. Hours later, the explosion propagates to the stellar surface, instigating a burst of photons across the electromagnetic (EM) spectrum with a total energy of around ${\sim} 10^{49}\,\mathrm{erg}$. 

The absolute majority of the supernovae have so far been observed only through their EM signature. While this provided a wealth of information about the supernovae, there is a limit to what photons alone can tell. This is because the EM burst forms at the outer edge of the star, so it contains limited information about the inner regions, where the CCSN central engine is located. 

Since gravitational waves are generated by the aspherical motion in the inner regions, they contain information about the dynamics of the central engine. In addition, neutrinos carry information about the thermodynamic conditions at the surface of the PNS. Therefore, observations of GWs and neutrinos will enable us to probe the central engine in a completely new way. In particular, this may allow us to extract information about the mechanism that produces the explosion. Moreover, such an observation may reveal the rotation and the structure of the innermost regions of the star.   

In this chapter, we summarize our current understanding of GW emission from CCSNe with a focus on established results. As we discuss below, most of what we know about CCSN GWs comes from sophisticated multi-dimensional simulations. Wherever possible, we will complement these results with simple back-of-the-envelope calculations to highlight the underlying physical principles. We also briefly discuss GW detection strategies focusing on the potential provided by a possible multi-messenger observation. Finally, we outline the main gaps and limitations in our current understanding of the subject. 

\section{Basic overall picture}
\label{sec:overall_picture}

In this section, we present a brief and basic overview of CCSNe and the associated GW emission mechanisms. 

\subsection{The road to core collapse}

Stars spend most of their lifetime in the main sequence phase, slowly burning their hydrogen fuel into helium. For a star with initial mass $M$, this phase lasts ${\sim} 10^{10} \, ( {M}/{M_\odot} )^{-2.5}$ years, where $M_\odot$ is the solar mass \cite{shapteu:83}. After exhausting hydrogen, the subsequent stages of nuclear burning proceed at much faster pace, producing heavier and more bound nuclei. Due to their lower temperatures, the least massive stars evolve into white dwarfs (WDs), as electron degeneracy pressure prevents the star from achieving sufficient densities and temperatures to ignite their carbon oxygen cores. The stars with initial mass $\lesssim 8{-}9 M_\odot$ evolve into carbon oxygen white dwarfs, while the stars with $8 M_\odot \lesssim M \lesssim 9 M_\odot$ evolve into oxygen-neon-magnesium (ONeMg) WDs. The exact mass limits that separate different evolutionary paths are not known. They depend on a number of free and poorly constrained parameters such as metallicity, mass loss, convection, and interaction with a binary companion \cite{snhandbook17}. 

Slightly more massive stars ($9 M_\odot \lesssim M \lesssim 10 M_\odot$) may also develop ONeMg cores, but instead of evolving into ONeMg WDs, they may produce supernovae. The electrons in the cores become degenerate and get captured by Ne and Mg nuclei. This depletes pressure and triggers collapse. Unless thermonuclear burning of O can stop the infall, the core collapses to a PNS, while the shock launched at bounce expels the stellar envelope and produces an {\emph{electron-capture supernova}} (ECSNe). Alternatively, if the thermonuclear burning is strong enough to overpower the collapse, the star may undergo a thermonuclear explosion and leave behind an iron-rich white dwarf \cite{snhandbook17}. 

For massive stars with initial masses $9 M_\odot \lesssim M \lesssim 100 M_\odot$, nuclear burning proceeds all the way beyond silicon burning, forming an iron core. At $T \sim 10^{10}\,\mathrm{K}$, as a result of collision with high-energy photons, iron nuclei dissociate into alpha particles and free nucleons. This absorbs thermal energy, leading to further contraction of the core. In addition, nuclei and free protons capture electrons, further reducing pressure. This triggers dynamical collapse of the core.

The stars with mass $\gtrsim 100M_\odot$ follow a different path. After central carbon burning, the temperature becomes high enough that photons start producing $e^+$ and $e^-$ pairs, converting thermal energy to rest mass. This reduces adiabatic index below $4/3$, which triggers gravitational collapse. This collapse is expected to proceed until BH formation for stars with mass $M \gtrsim 260M_\odot$. In stars in the intermediate mass range ($100 M_\odot - 200 M_\odot$) the thermonuclear reactions may be strong enough to overcome gravitational collapse and power \emph{pair-instability supernovae}, which might explode with energies as high as $\sim 10^{53}$ erg \cite{snhandbook17}.

\subsection{Core collapse and road to explosion}

Once triggered, the dynamical collapse of the core proceeds with an accelerating pace on a free-fall timescale $ {\sim} 0.3\,\mathrm{s}$. During collapse the core splits into two parts. The outer parts plunge supersonically, while the inner core descends with subsonic speed. Upon reaching supranuclear densities, nuclear matter stiffens, abruptly halting the collapse. The inner core bounces, launching a shock wave into the still-infalling outer core. As it progresses outward, the shock loses its energy to dissociation of iron nuclei, turning into a stalled accretion shock at ${\sim} 150\,\mathrm{km}$ within ${\sim} 10\,\mathrm{ms}$ after formation. In order to produce an explosion and leave behind a stable neutron star, the shock has to "revive" within a few hundred milliseconds and expel the infalling outer shells. Otherwise, the stellar matter keeps piling up on top the PNS, which eventually pushes the PNS beyond its stability limit, leading to BH formation.

How exactly the shock revives is a topic of an active research. According to our current understanding (recently reviewed by \cite{janka:12a,mueller:20review}), the explosion mechanism depends on a number of factors, with rotation perhaps being the most important differentiating factor. As the hot PNS cools and contracts, it emits ${\sim} 10^{53}\,\mathrm{erg}$ of its binding energy as neutrinos with ${\sim} 10^{52}\,\mathrm{erg/s}$ luminosity for ${\sim} 10\,\mathrm{s}$. Some of these neutrinos are absorbed behind the shock. In the so-called gain region, the neutrino heating exceeds cooling. This drives neutrino-driven hot-bubble convection. In addition, in some cases, the shock may become subject to the standing-accretion shock instability (SASI), which drives large-scale non-radial oscillations of the shock \cite{blondin:03}. When developed, these instabilities enhance neutrino heating, which energizes the shock to expel the stellar envelope and produce a supernova explosion with energies up to ${\sim} 10^{51}\,\mathrm{erg}$ energy. In rapidly rotating models, magnetic fields transfer the rotational kinetic energy of the PNS to the shock. For a millisecond period PNS, the rotational energy is ${\sim} 10^{52}\,\mathrm{erg}$, which may empower the supernovae with these energies (aka \emph{hypernovae}) and may even lead to long gamma-ray bursts (\eg \cite{janka:12a, mueller:20review} for recent reviews). 

While modern simulations of CCSNe confirm this baseline picture, there are still many open questions. The number of 3D simulations is rapidly growing, but a comprehensive understanding of the dependence of the explosion properties on model parameters (including progenitor mass and rotation as well as physical and numerical model assumptions) is still in its infancy. Although it is clear that convection and SASI help the explosion, we do not know how strongly these instabilities develop across different models. For example, there are models that produce explosions but do not exhibit any SASI (e.g., \cite{burrows:20}). As a result, despite all the impressive progress, it is not entirely clear which models explode (and if so, how strongly) and which models do not. Due to the close connection between the aspherical motion behind the shock, which is the main source of the GW emission, and the explosion dynamics, there is a potential for probing explosion mechanism of CCSNe using the GWs. A more detailed discussion of the explosion mechanism of CCSNe is beyond the scope of this chapter (see \cite{mueller:20review} for a recent review and references therein). 

\subsection{Generation of gravitational waves}

The GW strain $h$ emitted by a source at a distance $D$ can be approximately estimated using the quadrupole formula \cite{shapteu:83}
\begin{equation}
    h= \frac{2 G}{D c^4} \frac{d^2 Q}{dt^2},
\end{equation}
where $G$ is the gravitational constant, $c$ is the speed of light, and ${Q}$ is the mass quadrupole moment of the system. To an order of magnitude, for an object of mass $M$ undergoing quadrupole motion with speed $\upsilon$, 
\begin{equation}
\label{eq:h_est}
h D \sim \epsilon r_\mathrm{Sch} \frac{\upsilon^2}{c^2}, 
\end{equation}
where $\epsilon$ is a measure of non-sphericity of the source ($=0$ for spherical source) and $r_\mathrm{Sch}=GM/c^2$ is the Schwarzschild radius associated with the mass $M$. The GW luminosity can be estimated as \cite{shapteu:83}:
\begin{equation}
    \frac{dE_\mathrm{GW}}{dt} \sim \epsilon^2 \frac{c^5}{G} \left(\frac{r_\mathrm{Sh}}{R}\right)^2
    \left(\frac{\upsilon}{c}\right)^6,
\end{equation}
where $R$ is a characteristic size of the source. Thus, in order to produce strong GWs, we need a compact object with fast aspherical motion. For CCSNe, this means that the PNS should produce most of the GWs. This intuition is supported by modern CCSN simulations: while the asymmetric flow outside the PNS does produce some GWs, the PNS dynamics is responsible for most of the emitted GWs. Assuming $\epsilon=0.1$, $\upsilon/c \sim 0.1$, and total emission time of ${\sim} 1\,\mathrm{s}$, we obtain crude upper limits for the GW strain $hD \sim 300\,\mathrm{cm}$ and the total emitted energy $E_\mathrm{GW} \sim 10^{47}\,\mathrm{erg}$ ($\sim 10^{-7}M_\odot c^2$) \cite{kotake:17}. 

The dynamics of PNS in rapidly rotating models is vastly different from that in slowly or non-rotating models. In rapidly rotating models the centrifugal force leads to slower bounce along equatorial plane than along the rotation axis. As a result, the PNS is born with oblate perturbation. This triggers PNS oscillations that last for ${\sim} 10{-}20\,\mathrm{ms}$ \cite{richers:17}. In addition, the PNS may be subject to rotational non-axisymmetric instabilities, which deform the PNS into a rotating non-axisymmetric shape. Rotation of non-axisymmetric object produces long-lasting GW emission \cite{kotake:13review}. 

In slowly or non-rotating models, the centrifugal force has little dynamical impact. Instead, convection and SASI perturb the PNS and excite its oscillations \cite{murphy:09}. In the following, we discuss these two cases separately. 

Furthermore, there are sub-dominant (and/or somewhat exotic) mechanisms for generating GWs: besides perturbing the PNS, convection and SASI can directly emit GWs. The variations in neutrino luminosity in the different regions produce anisotropic flux of neutrinos. The dense matter inside PNS may undergo phase transition, which may lead to a "mini" second core-collapse of the PNS. Finally, if the PNS accumulates more mass than it can support, it collapses to a BH. Each of these processes can emit GWs and we will discuss them in more detail later.

\section{Non-rotating and slowly-rotating case}
\label{sec:slow_rotating}

\begin{figure}
\begin{center}
 \includegraphics[angle=0,width=1\columnwidth,clip=false]{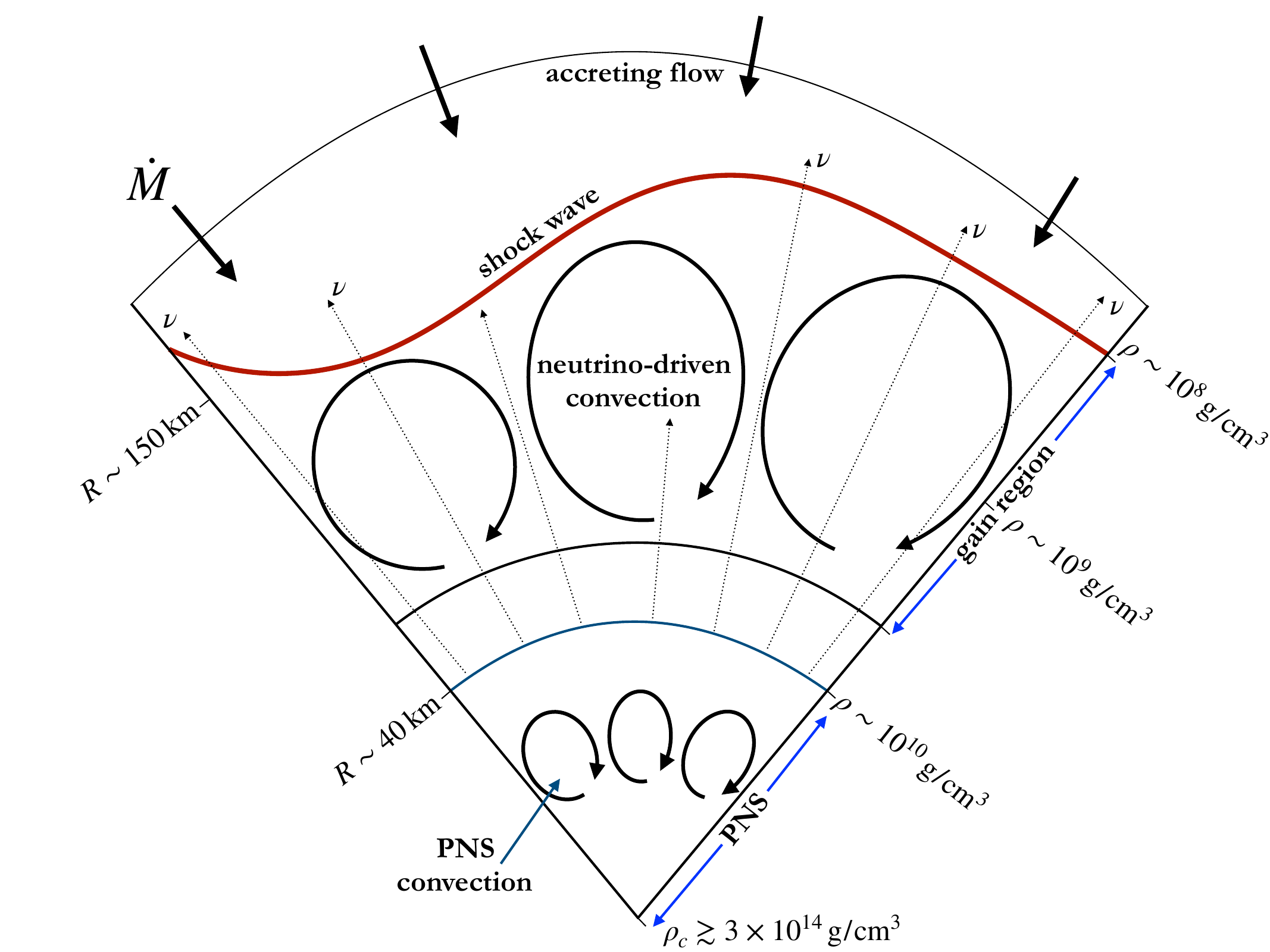}
  \caption{The schematic depiction of the CCSN central engine for slowly rotating case. The neutrinos emitted by the proto-neutron star (PNS) drive neutrino-driven convection in the gain region. The diffusion of neutrinos out of the PNS leads to negative radial gradient of the lepton number, driving PNS convection. The SASI drives large-scale oscillations of the shock with low $\ell$ number. \label{fig:scheme}}
\end{center}
\end{figure}

\begin{figure}
\begin{center}
 \includegraphics[angle=0,width=0.75\columnwidth,clip=false]{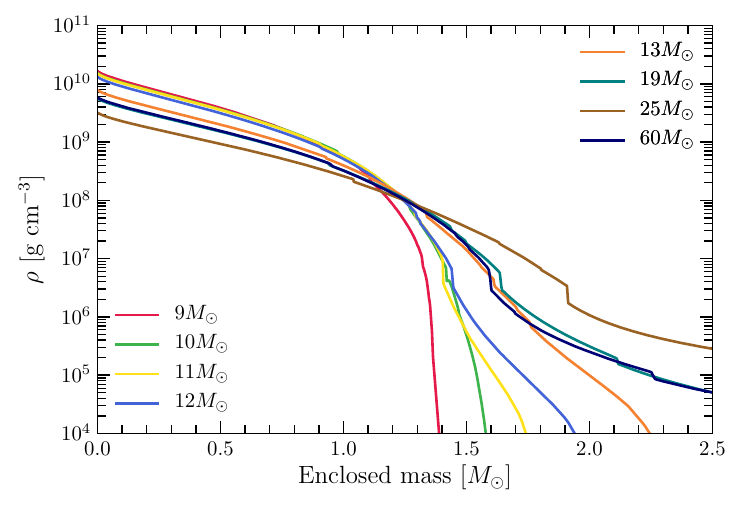}
  \caption{The radial profile of density of the progenitor models studied by D.~Radice, V.~Morozova, A.~Burrows, D.~Vartanyan, H.~Nagakura, \textit{Characterizing the Gravitational Wave Signal from Core-Collapse Supernovae}, Astrophys.J.Lett. \textbf{876}, L9 (2019) \cite{radice:19gw}. \label{fig:progenitors_radice19}}
\end{center}
\end{figure}

\begin{figure}
\begin{center}
 \includegraphics[angle=0,width=0.75\columnwidth,clip=false]{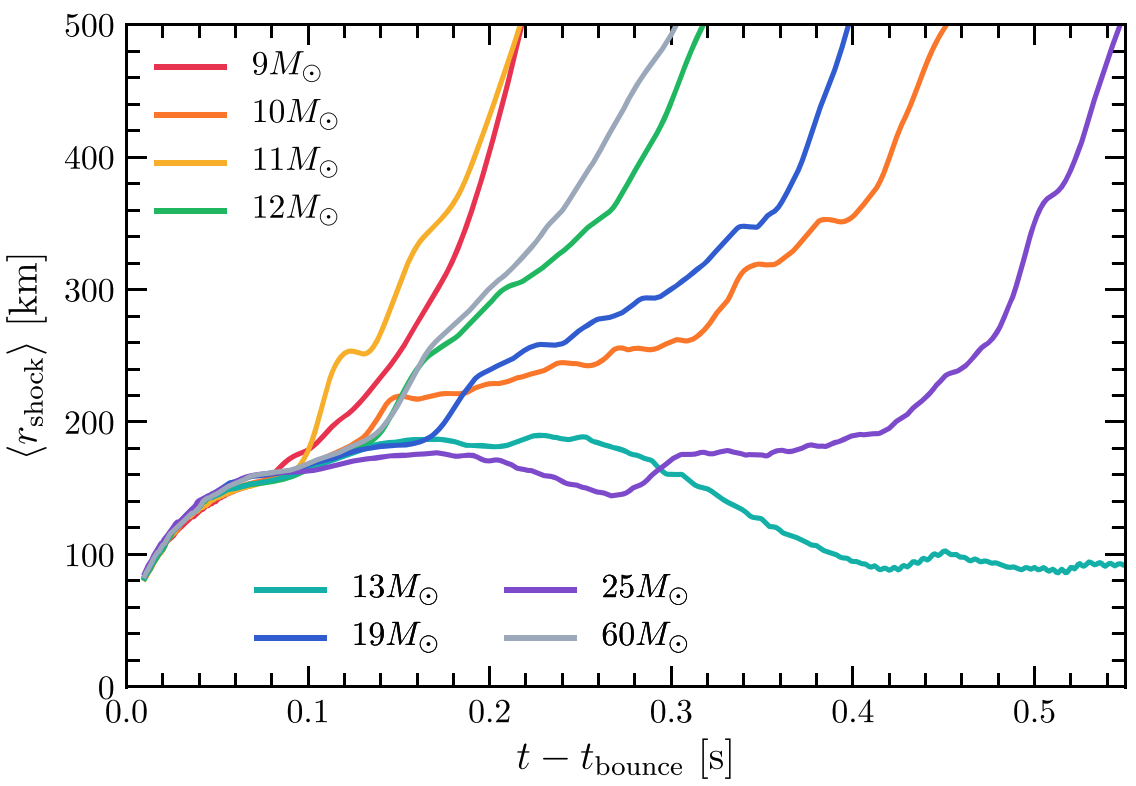}
  \caption{The shock radius as a function of time for models. Reprinted from D.~Radice, V.~Morozova, A.~Burrows, D.~Vartanyan, H.~Nagakura, \textit{Characterizing the Gravitational Wave Signal from Core-Collapse Supernovae}, Astrophys.J.Lett. \textbf{876}, L9 (2019) \cite{radice:19gw}. \textcopyright~AAS. Reproduced with permission.  \label{fig:rshock_vs_t_radice19}}
\end{center}
\end{figure}

\begin{figure}
\begin{center}
 \includegraphics[angle=0,width=0.75\columnwidth,clip=false]{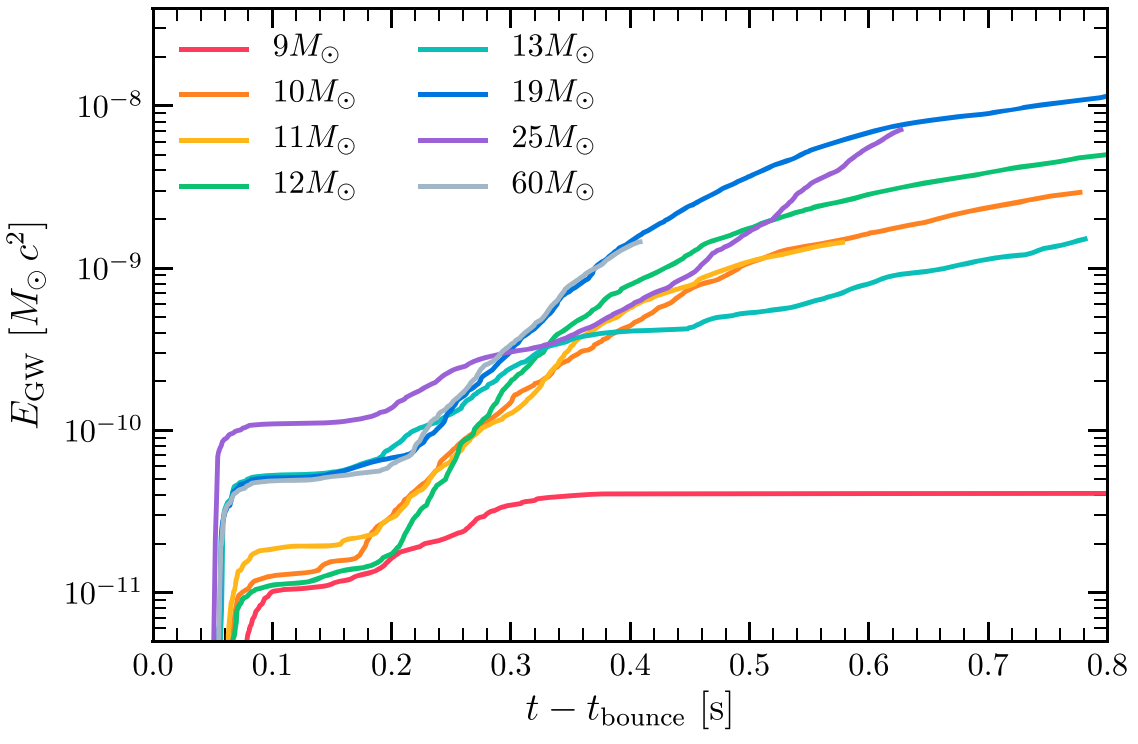}
  \caption{GW energy as a function of time. Reprinted from D.~Radice, V.~Morozova, A.~Burrows, D.~Vartanyan, H.~Nagakura, \textit{Characterizing the Gravitational Wave Signal from Core-Collapse Supernovae}, Astrophys.J.Lett. \textbf{876}, L9 (2019) \cite{radice:19gw}. \textcopyright~AAS. Reproduced with permission.  \label{fig:egw_vs_t_radice19}}
\end{center}
\end{figure}

Most stars are expected to rotate slowly (e.g., \cite{deheuvels:14}). As mentioned above, such stars remain quasi-spherical during bounce and early ($\lesssim 10\,\mathrm{ms}$) post-bounce phase without emitting any GWs. Instead, the multi-dimensional hydrodynamic flows that develop later lead to GW emission. 

The shock wave, launched at bounce, leaves behind a negative gradient of entropy while traveling outward within the first few milliseconds. This triggers prompt convection in the post-shock region ${\sim} 10\,\mathrm{ms}$ after bounce. Prompt convection lasts for ${\lesssim} 50\,\mathrm{ms}$ and is followed by a quiescent period of ${\lesssim} 100\,\mathrm{ms}$. Afterward, the neutrino heating in the gain region drives the neutrino-driven convection behind the shock, while in some models the shock may develop SASI oscillations. The neutrino emission from the PNS creates negative gradient of lepton number in the region $10\lesssim r \lesssim 25 \,\mathrm{km}$, driving vigorous {\emph{PNS convection}} for few seconds after formation. These instabilities, schematically depicted in Fig.~\ref{fig:scheme}, perturb the PNS and excite its oscillations, which then produce GWs. Once the system starts to transition to explosion, asymmetric shock expansion generates a non-oscillatory shift in the GW strain, while any material that may fall back onto PNS generates additional high-frequency GWs. Since convection and SASI develop from stochastic perturbations, the resulting dynamics and GW signal are expected to have a stochastic time variability.

The impact of these instabilities varies from model to model, leading to distinct GW signatures. For this reason, in order to cover a wide range of scenarios, below we discuss eight different models with (initial) progenitor masses ranging from $9M_\odot$ to $60M_\odot$ that were recently simulated by \cite{radice:19gw,burrows:20} with sophisticated 3D neutrino-hydrodynamics and microphysics\footnote{Here, we focus on results only from 3D simulations of CCSNe. In 2D axisymmetric simulations, energy in large scale motion is artificially large due to inverse energy cascade of 2D turbulence (e.g., \cite{mueller:20review} for a recent review). Also, SASI is restricted to axisymmetric modes in 2D. These effects lead to inaccurate predictions of GW signal.}. Figure~\ref{fig:progenitors_radice19} shows the radial density profiles of these progenitors. This reveals that the progenitors have vastly different structures, ranging from models with little mass above iron core (e.g. $9M_\odot$ model) to models with significant mass above the core (e.g. $25M_\odot$ model). This leads to a rich variety of outcomes.

\begin{figure}
\begin{center}
 \includegraphics[angle=0,width=1.0\columnwidth,clip=false]{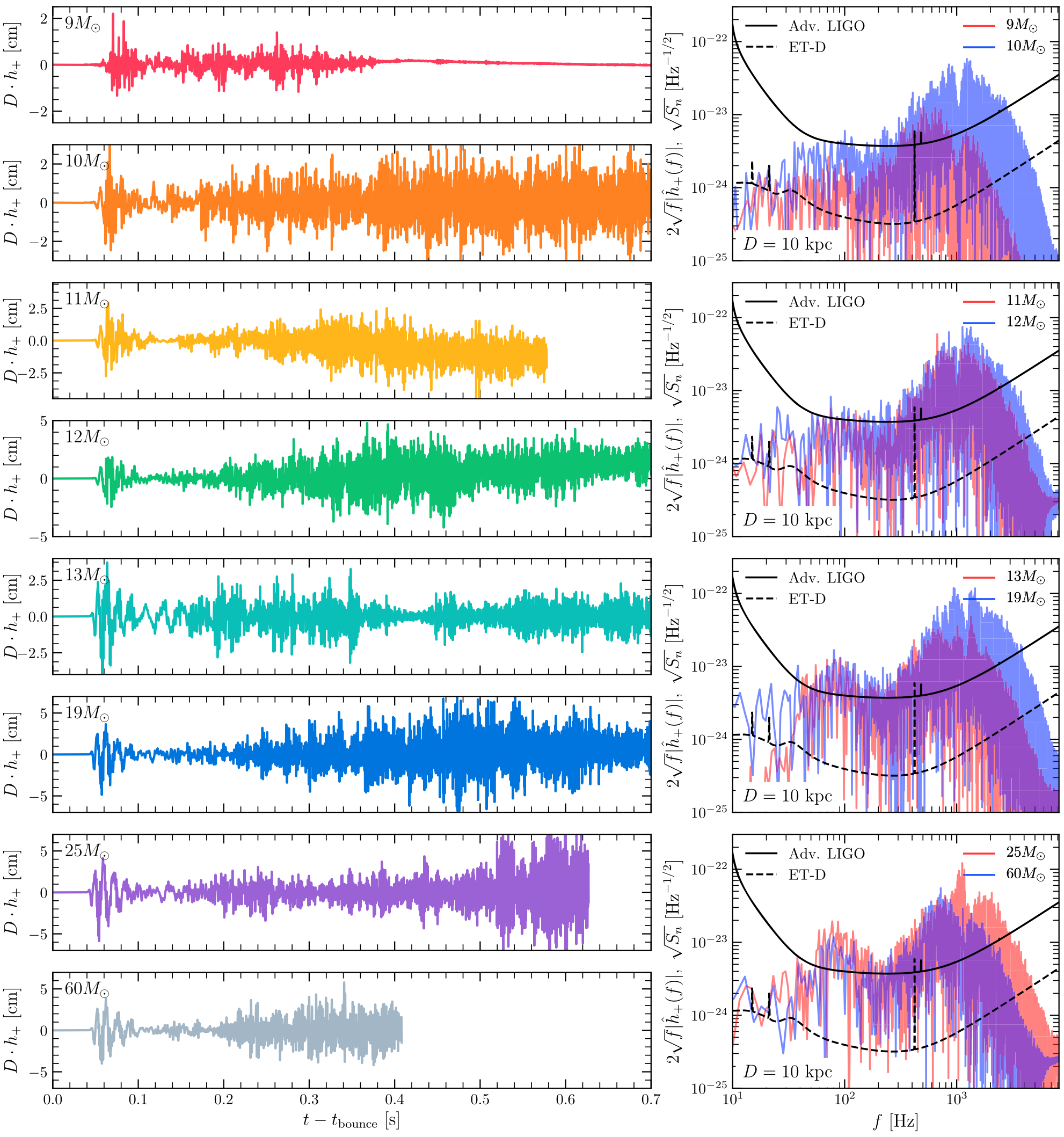}
  \caption{{\bf Left panels:} GW strain as a function of time for eight models of \cite{radice:19gw}. {\bf Right panels:} The spectral energy distribution of the GW signal together with the sensitivity curves of Advanced LIGO and ET-D detectors for sources at $10\,\mathrm{kpc}$ distance. Reprinted from D.~Radice, V.~Morozova, A.~Burrows, D.~Vartanyan, H.~Nagakura, \textit{Characterizing the Gravitational Wave Signal from Core-Collapse Supernovae}, Astrophys.J.Lett. \textbf{876}, L9 (2019) \cite{radice:19gw}. \textcopyright~AAS. Reproduced with permission.  \label{fig:h_vs_t_radice19}}
\end{center}
\end{figure}

As we can see in the evolution of the average shock radius versus time, shown in Fig.~\ref{fig:rshock_vs_t_radice19}, all of the models, except the $13M_\odot$ model, transition to explosion within $200{-}500\,\mathrm{ms}$ after bounce. All models exhibit strong neutrino-driven convection, but only $13M_\odot$ and $25M_\odot$ models develop SASI. Fig.~\ref{fig:egw_vs_t_radice19} shows the energy of emitted GWs as a function of time. Comparing this to the shock radius evolution, we can see that there is no strong correlation between the shock radius and the GW energy. For example, the $9M_\odot$ model, which undergoes fast transition to explosion, radiates the smallest amount of energy in GWs (${\sim} 4\times 10^{-11} M_\odot c^2\simeq 7.2\times10^{43}\,\mathrm{erg}$) among these models. This is because in this model, as well as in other low-mass CCSN (and ECSN) progenitors, little mass is available outside of the iron core (cf. Fig.~\ref{fig:progenitors_radice19}). The shock and post-shock flow remains quasi-spherical in the post-bounce phase. As a result, little GW emission takes place, especially beyond ${\sim} 300\,\mathrm{ms}$ after bounce, as we can see in the plot of GW strain versus time, shown on left panels of Fig.~\ref{fig:h_vs_t_radice19}. On the other hand, the $13M_\odot$ model, which does not exhibit explosion, experiences strong convection and SASI. The plot of GW strain versus time (cf. Fig.~\ref{fig:h_vs_t_radice19}) reveals strong high-frequency GW emission throughout post-bounce evolution within the time span indicated in this figure. This leads to strong GW emission with $\gtrsim 10^{-9} M_\odot c^2 \simeq 2\times10^{45}\,\mathrm{erg}$ energy. 

Assuming a distance of 10 kpc to the source and a perfect knowledge of waveform, the spectral energy density of GWs is shown on the right panels of Fig.~\ref{fig:h_vs_t_radice19} together with the sensitivity curves of the Advanced LIGO and the proposed Einstein telescope in the D configuration. The corresponding optimal SNR for Advanced LIGO ranges from ${\sim} 1$ for the $9M_\odot$ model to ${\sim} 10$ for the $19M_\odot$ model. For the third-generation detector, the corresponding SNRs range from ${\sim} 10$ to ${\sim} 100$. This suggests that third-generation detectors are necessary for a more reliable detection of GWs from CCSNe across our galaxy.  

Below, we discuss each of the GW emission mechanisms in more detail and identify the physical mechanisms responsible for them. 

\subsection{Early Quasi-Periodic Signal}
\label{sec:hydro.prompt}

All models experience the early quasi-periodic signal that develops between $ \sim 10\,\mathrm{ms}$ and $\sim 50\,\mathrm{ms}$ after bounce, as can be seen in the plot of GW strain as a function of time shown in Fig.~\ref{fig:h_vs_t_radice19}. It is also visible as a track with ${\sim} 100\,\mathrm{Hz}$ frequency in the GW spectrogram shown in Fig.~\ref{fig:gw.specgram} for the $25M_\odot$ model. The origin of this signal has been ascribed to a number of processes, including the prompt post-shock convection. The contributions of the shock oscillations \cite{yakunin:10} and acoustic waves in the post-shock region have also been emphasized \cite{mueller:13}. Depending on model, the GW strain of the early quasi-periodic signal can reach up to $hD\sim 2-5 \,\mathrm{cm}$, but due to their short duration, their contribution to the overall signal is somewhat modest (cf. Fig.~\ref{fig:egw_vs_t_radice19}).

\subsection{PNS convection}
\label{sec:pns_convection}

The PNS convection has been invoked as a mechanism for exciting PNS oscillations and generating GWs \cite{andresen:17}. However, the $9M_\odot$ model, despite exhibiting vigorous convection throughout its evolution \cite{nagakura:20}, emits little GWs in the wake of runaway shock expansion at ${\sim} 300\,\mathrm{ms}$ after bounce (cf. Fig.~\ref{fig:h_vs_t_radice19}). This means that the PNS convection does not have significant impact on the GW emission, at least for this model. {In the rest of the models, it is hard to make a definite statement because the neutrino-driven convection and SASI overlap in time with the PNS convection, at least within the time span covered by the simulations.} That said, evidence of stronger contribution of the PNS convection to the GW signal was suggested in \cite{andresen:17, mezzacappa:20}. This discrepancy points to potential model dependency of the PNS convection, which call for more detailed studies \cite{powell:19}. 

\subsection{Neutrino-Driven Convection}
\label{sec:hydro.driven}

As mentioned above, neutrino heating drives neutrino-driven convection in the post-shock region (cf. Fig~\ref{fig:scheme}). Since this occurs in an accreting flow, in order to develop the convection has to grow faster than the advection timescale. Otherwise, it will be accreted out from the gain region before it can develop. This condition can be formulated in terms of the parameter
\begin{equation}
    \chi = \int_{R_\mathrm{g}}^{R_\mathrm{sh}} \frac{\mathrm{Im}\, N }{|\upsilon_r|} dr,
\end{equation}
where $\mathrm{Im}\,N$ is the imaginary part of the Brunt-V\"ais\"al\"a frequency $N$. Convection was found to develop for $\chi\gtrsim 3$ \cite{foglizzo:06}. The size of the largest eddies is determined by the width of the gain region: a convective region with height $H=R_\mathrm{sh}-R_\mathrm{g}$ can fit $\ell$ pairs of circular vortices of size $H$ along the average circumference of the region. This results in
\begin{equation}
    \ell \sim \frac{\pi}{2} \frac{R_\mathrm{sh}+R_\mathrm{g}}{R_\mathrm{sh}-R_\mathrm{g}}.
\end{equation}
This shows that $\ell$ is smaller for larger shock radii. When shock is small, eddies with $\ell$ in range $4-8$ have been observed. When shock is large and expanding, large-scale modes with $\ell=1$ and $\ell=2$ become dominant \cite{mueller:20review}.

The contribution of the GWs emitted directly by neutrino-driven convection to the overall signal from CCSNe is somewhat minor. However, as mentioned earlier, convection creates funnels of accretion onto PNS, which excites PNS oscillations that generate stronger GWs.

\subsection{SASI}
\label{sec:hydro.sasi}

SASI represents non-radial oscillations of the shock. It develops due to the so-called advective-acoustic cycle. Any perturbation of the shock generates vorticity waves in the post-shock region, which are then advected by the flow toward the PNS. As it approaches the PNS, it encounters steep gradient of density. The vorticity wave distorts the isodensity surfaces of the PNS, producing pressure perturbations and generating acoustic waves. These acoustic waves travel outward and reach the shock. This perturbs the shock more, generating new vorticity and further amplifying the oscillations of the shock \cite{foglizzo:07}\footnote{A purely acoustic mechanism has also been considered \cite{mezzacappa:20}, but latest detailed analyses favor the advective-acoustic mechanism for SASI. See, e.g., \cite{mueller:20review} for a recent review.}.

\begin{figure}
    \centering
    \includegraphics[width=4in]{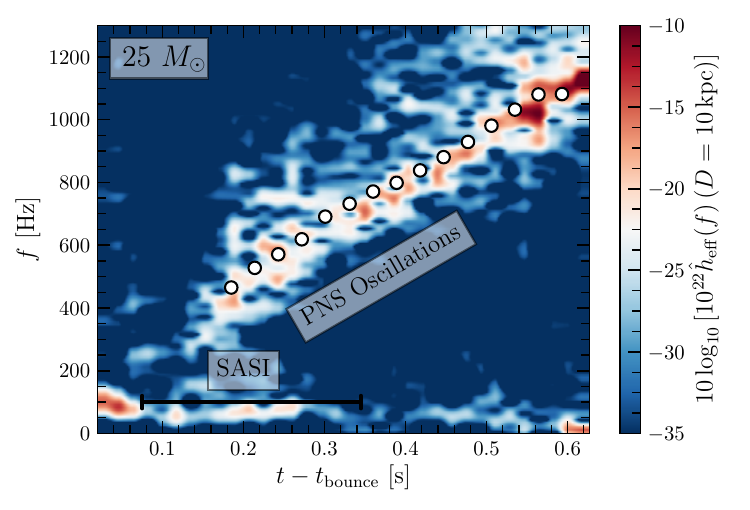}
    \caption{Spectrogram of the GW signal for the $25 M_\odot$ model. The white dots represent the eigenfrequencies of the quadrupolar $f$- and $n = 1, 2$, $g$-modes of the PNS obtained from the linear perturbation analysis. Reprinted from D.~Radice, V.~Morozova, A.~Burrows, D.~Vartanyan, H.~Nagakura, \textit{Characterizing the Gravitational Wave Signal from Core-Collapse Supernovae}, Astrophys.J.Lett. \textbf{876}, L9 (2019) \cite{radice:19gw}. \textcopyright~AAS. Reproduced with permission.}
    \label{fig:gw.specgram}
\end{figure}

Due to the advective-acoustic cycle, one SASI period is a sum of advection and sound-crossing time between the shock and PNS. Since post-shock flow is subsonic, the advection timescale is much larger than the sound-crossing time. Using this, we can obtain an approximate estimate for the SASI period
\begin{equation}
    T_\mathrm{SASI} \sim \int_{R_\mathrm{PNS}}^{R_\mathrm{sh}} 
    \frac{dr}{|\upsilon_\mathrm{r}|}.
\end{equation}
The advection velocity scales as $\upsilon_r \sim \upsilon_\mathrm{PS} r/R_\mathrm{PS}$, where $\upsilon_\mathrm{PS} \sim \sqrt{G M_\mathrm{PS}/R_\mathrm{S}}/7$ is the post-shock velocity \cite{mueller:20review}. This yields
\begin{equation}
    T_\mathrm{SASI} \sim \frac{R_\mathrm{sh}}{\upsilon_\mathrm{PS}} \ln\left(\frac{R_\mathrm{sh}}{R_\mathrm{PNS}}\right) \sim 20 \,\mathrm{ms} \left(\frac{R_\mathrm{sh}}{100\,\mathrm{km}}\right)^{3/2} \ln\left(\frac{R_\mathrm{sh}}{R_\mathrm{PNS}}\right).
\end{equation}
This estimate is consistent with the frequency we observe in the spectrogram of the GW signal shown in Fig.~\ref{fig:gw.specgram}. It also shows that the frequency of the GWs from SASI should decrease with increasing shock radius. This decrease can be identified in Fig.~\ref{fig:gw.specgram} as the shock starts to expand at around $\sim 400\, \mathrm{ms}$ after bounce. Further expansion of the shock leads to diminishing of the SASI activity as the advective-acoustic cycle responsible for SASI cannot keep up with rapidly expanding shock. 

Using Eq.~(\ref{eq:h_est}), and assuming gain region mass of $10^{-2}M_\odot$, $\upsilon/c\sim 0.1$, and $\epsilon=0.1$, we obtain $hD\sim 1\,\mathrm{cm}$, which is consistent with the results of numerical simulations \cite{mezzacappa:20}. In addition, similarly to neutrino-driven convection, SASI modulates accretion onto PNS, which excites PNS oscillations and generates more GWs.   

\subsection{Protoneutron Star Pulsations}
\label{sec:pns}

Despite the continued accretion and the presence of hydrodynamics instabilities, the inner region of a CCSN, encompassing the stalled shock and the PNS, can be considered as stationary with good precision, \eg~\cite{janka:01}. In particular, the density stratification is close to hydrostatic behind the shock wave during the shock stagnation phase, and at all times in the PNS. For this reason, the PNS oscillations can be described using linear perturbation theory with good accuracy. 

In the linear theory, the Lagrangian displacement field $\pmb{\xi}(t,\pmb{r})$, describing the deformation of the PNS, obeys the equation of motion
\begin{equation}\label{eq:linear.theory}
    \frac{\partial^2 \pmb{\xi}}{\partial t^2} + \pmb{C}\pmb{\xi} = \pmb{F}\,,
\end{equation}
where $\pmb{F}$ is the perturbing force and $\pmb{C}$ is a differential operator. To illustrate the basic ideas, we consider the non-rotating, spherical, Newtonian case. We also restrict ourselves to adiabatic perturbations, which means that we neglect heat and compositional changes in the fluid elements as they are displaced. In this case, $\pmb{C}$ is a self-adjoint operator. Its exact expression can be found in \eg~\cite{friedman:78}, which also discusses the more general case of rotating stars. For our purposes it is only important to emphasize that $\pmb{C}$ depends only on the unperturbed density, pressure, and composition of the PNS.
See Ref.~\cite{kokkotas:99} for a review of the general-relativistic formalism and Ref.~\cite{bruenn:96} for a discussion on the impact of non-adiabatic effects.

In the case we are considering, it is possible to show that the eigenfunctions associated with the time-independent eigenvalue problem
\begin{equation}\label{eq:normal.modes}
    \pmb{C} \pmb{\xi}_\alpha(\pmb{r}) = \omega^2_\alpha \pmb{\xi}(\pmb{r}),
\end{equation}
with appropriate boundary conditions (more below), form a complete set. They can be normalized as
\begin{equation}
    \int \pmb{\xi}_{\alpha'}^\ast(\pmb{r})\cdot \pmb{\xi}_{\alpha}(\pmb{r}) \rho(\pmb{r}) \dd^3 \pmb{r} =
    M R^2 \delta_{\alpha' \alpha},
\end{equation}
where $M$ and $R$ are the total mass and radius of the background solution. This means that any pulsation pattern of the star can be decomposed as
\begin{equation}
    \pmb{\xi}(t,\pmb{r}) = \sum_{\alpha} A_\alpha(t) \pmb{\xi}_\alpha(\pmb{r}).
\end{equation}
Substituting this expansion into \eqref{eq:linear.theory} we find the evolution equation for the amplitudes
\begin{equation}
    \frac{\dd^2 A_\alpha(t)}{\dd t^2} + \omega_\alpha^2 A_\alpha(t) =
    \frac{1}{M R^2} \int \pmb{\xi}^\ast_\alpha(\pmb{r}) \cdot \pmb{F}(t,\pmb{r}) \rho(\pmb{r})  \dd^3 \pmb{r}.
\end{equation}
In other words, linear perturbation theory shows that the PNS oscillations can be written as superposition of normal modes, each endowed with its characteristic frequency, and that the normal modes behave like a system of uncoupled forced harmonic oscillators. An important consequence is that the GW spectrum from CCSNe is predicted to contain discrete frequencies, related to the eigenvalues of the operator $\pmb{C}$, that depend only on the structure of the PNS and not on the nature of the forcing. This suggests that GWs from CCSNe could be used to probe the interior structure of the PNS as it cools and contracts. {It is important to remark that the decomposition in normal modes is possible even in the presence of rotation. However, this decomposition is not strictly valid in the GR formalism, in which the modes lose energy due to GW emission (quasi-normal modes) and do not form a complete basis~\cite{kokkotas:99}.}

Some of the first studies of the normal modes of PNSs with realistic temperature and compositional profiles were presented by Refs.~\cite{ferrari:03, sotani:16, torres-forne:19}. With the availability of long-term 3D CCSN simulations with realistic microphysics, it is now also possible to validate perturbation theory with nonlinear simulations \cite{radice:19gw, powell:20, sotani:20}. The typical starting point for the perturbation theory is to decompose the Lagrangian displacement in vector spherical harmonics:
\begin{equation}
    \pmb{\xi} = \left[ \xi^r \hat{\pmb{e}}_r + \xi^\perp \left(
        \hat{\pmb{e}}_\theta \frac{\partial}{\partial \theta} + \hat{\pmb{e}}_\varphi \frac{1}{\sin\theta}\frac{\partial}{\partial \varphi} \right) \right] Y_l^m(\theta,\varphi).
\end{equation}
Substituting this expression into Eq.~\eqref{eq:normal.modes} yields a set of decoupled nonlinear one-dimensional eigenvalue problems, one for each spherical harmonic $l, m$. The formulas are particularly simple in the Cowling approximation, that is when perturbations on the gravitational potential are neglected. This is a good approximation especially for high-order modes, and it is useful to illustrate the key ideas. With this approximation Eq.\eqref{eq:normal.modes} reduces to \cite{saio:93}
\begin{equation}\label{eq:cowling}
    \begin{aligned}
    &\frac{1}{r^2} \frac{\dd (r^2 \xi^r)}{\dd r} - \frac{g}{c_s^2} \xi^r +
       \left( 1 - \frac{L_l^2}{\omega^2} \right) \frac{\delta\! p}{\rho c_s^2} = 0, \\
    &\frac{1}{\rho}\frac{\dd \delta\!p}{\dd r} + \frac{g}{\rho c_s^2} \delta\! p + (N^2 - \omega^2) \xi^r = 0,
    \end{aligned}
\end{equation}
where $g$, $c_s$, $p$, $\rho$ are the local gravitational acceleration, sound speed, pressure, and density of the background star, $\delta\! p$ is the Eulerian pressure perturbations
\begin{equation}
    \delta\! p = - \gamma p \nabla \cdot \pmb{\xi} - \pmb{\xi}\cdot \nabla p, 
\end{equation}
$\gamma = (\partial \ln p/\partial \ln \rho)_{S, Y_e}$ being the adiabatic index, and $L_l$ and $N$ are the Lamb and Brunt-V\"ais\"al\"a frequencies defined as
\begin{equation}
    L_l^2 = \frac{l(l+1) c_s^2}{r^2}, \quad
    N^2 = g \left( \frac{1}{\gamma} \frac{\dd \ln p}{\dd r} - \frac{\dd \log \rho}{\dd r} \right).
\end{equation}
These equations are typically closed with regularity boundary conditions at the center ($\xi^r = 0$) and vanishing pressure perturbation at the surface of the PNS ($\delta\! p = 0$), see \cite{sotani:19} for a discussion of possible boundary conditions. These boundary conditions ensure that the operator $\pmb{C}$ is self-adjoint and that the resulting modes form a complete orthogonal basis.

We can gain some physical intuition on these equations by solving them in the short wavelength limit (WKB approximation) in which
\begin{equation}
    \xi^r, \delta\! p \sim \exp(i k_r r).
\end{equation}
Substituting this expression into Eq.~\eqref{eq:cowling} we obtain the (approximate) dispersion relation
\begin{equation}\label{eq:cowling.dispersion}
    k_r^2 \simeq \frac{(\omega^2 - L_l^2) (\omega^2 - N^2)}{\omega^2 c_s^2}.
\end{equation}
Oscillations can propagate in the radial direction if $k_r$ is real, otherwise the waves are exponentially damped (\emph{evanescent}). There are two possibilities:
\begin{enumerate}
    \item $\omega^2 > L_l^2, N^2$: these are called $p$-modes, because their restoring force is pressure;
    \item $\omega^2 < L_l^2, N^2$: these are called $g$-modes, because their restoring force is gravity.
\end{enumerate}
For large frequencies $\omega \gg 1$, $k_r \sim \frac{\omega}{c_s}$ which shows that higher-order $p$-modes have higher oscillation frequencies than lower order $p$-modes. Conversely, for small frequencies $\omega \ll 1$, $k_r \sim l(l+1) N^2/(\omega^2 r^2)$, which shows that higher-order $g$-modes, instead, have lower oscillation frequencies than lower-order $g$-modes. The frequencies of the $p$-modes are typically larger than that of the $g$-modes. The two families of modes are separated by the so-called $f$-mode, which is a mode with no radial nodes ($\xi^r, \delta\! p \neq 0$ for $0 < r < R$). If rotation is present, modes whose restoring force is the Coriolis force are also present. These are the so-called $r$-modes. Finally, in GR a new family of modes, the so-called $w$-modes, is also present. These are modes with characteristic frequencies of several kHz that originate from oscillations in the spacetime metric that have no classical counterpart \cite{kokkotas:99}. It is important to emphasize that, in the case of radial oscillations ($l = 0$), only the $p$-modes are present, since the Lamb frequency is zero.

\begin{figure}
    \centering
    \includegraphics[width=4.5in]{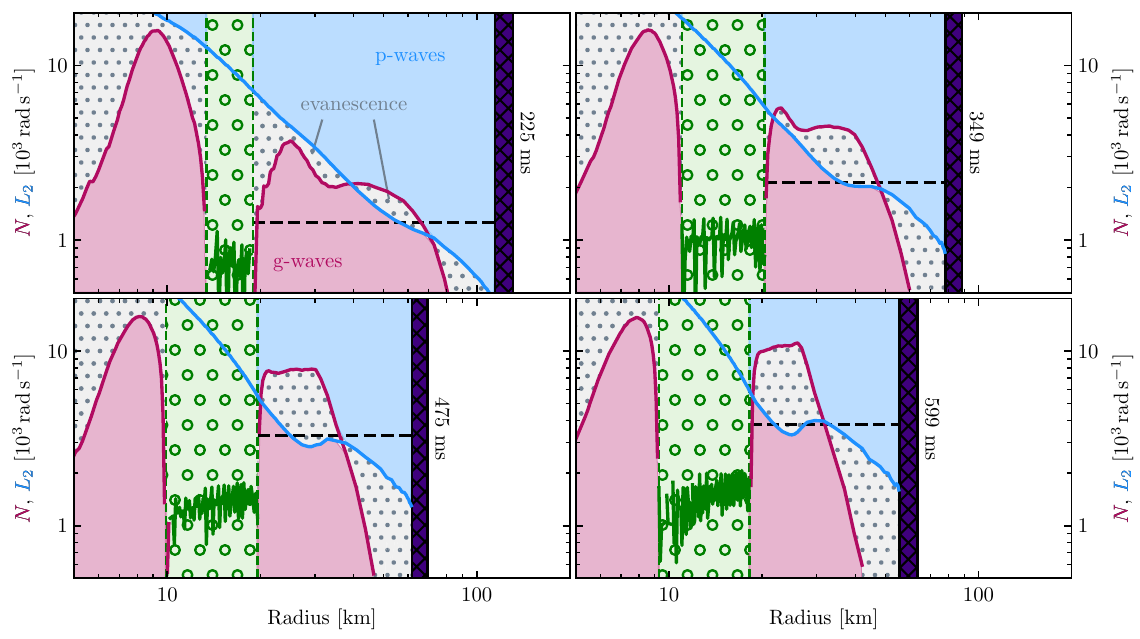}
    \caption{Propagation diagram for $p$ and $g$ modes in a contracting PNS. The solid regions represents regions where the modes can propagate. The vertical black line denotes the position of the shock. There are two regions where modes propagate separated by a convective region where $N^2 < 0$ ($|N|$ is shown as a green line there). Adapted from Gossan, Fuller, Roberts, \textit{Wave heating from proto-neutron star convection and the core-collapse supernova explosion mechanism}, MNRAS \textbf{491} 5376 (2020) \cite{gossan:20}. \textcopyright~Gossan et al. Reproduced with permission.}
    \label{fig:gossan}
\end{figure}

Figure \ref{fig:gossan} shows the characteristic evolution of $N$ and $L_2$ for a PNS taken from \cite{gossan:20}. These calculations are for a model simulated in 1D (assuming spherical symmetry) for which the shock stalls and explosion is not successful. However, the structure of the PNS depends only weakly on the outcome of the supernova explosion on the timescales relevant for the GW emission, so this model is representative. It is possible to see that there are two regions where $g$- and $p$-modes can propagate: one close to the surface, and one in the inner core of the PNS. They are separated by a region that is unstable to convection, corresponding to the PNS convection region \cite{gossan:20}. Because of the presence of this convective layer, the $g$-modes of the inner core are trapped and cannot be excited by perturbations acting on the surface of the star. It is important to remark that the propagation diagram offers only a partial description of the full spectrum of oscillations, since the WKB approximation is only valid for large mode wavenumbers. A full calculation is needed to determine the full spectrum of modes from PNSs.

Ab initio CCSN simulations are in good agreement with predictions from linear theory. Simulations show that most of the GW energy is radiated at a frequency coincident with that of a quadrupolar low-order $g$- or the $l = 2$ $f$-mode of the PNS \cite{murphy:09, mueller:13, andresen:17, morozova:18, radice:19gw}. Additionally, accretion modulated by the SASI, if present, is found to also excite quadrupolar oscillation modes in the PNS and contribute to a low frequency component of the signal \citep{andresen:17, radice:19gw}. 

Figure \ref{fig:gw.specgram} shows the GW signal of the $25\, M_\odot$ of \cite{radice:19gw}. The signal is characterized by a dominant feature that evolves with time and as the PNS contract. It evolves from ${\sim} 400\,\mathrm{Hz}$ at ${\sim} 150\,\mathrm{ms}$ to ${\sim} 1,100\,\mathrm{Hz}$ at ${\sim} 600\,\mathrm{ms}$ after bounce. This corresponds to the eigenfrequency of the $f$-mode of the PNS. Other models considered here exhibit a similar behavior. 

\subsection{Explosion phase signal}
\label{sec:late_signal}

Aspherical shock expansion during explosion represents anisotropic flow of matter. This produces non-oscillatory, slowly-varying changes of the quadrupole moment. The resulting GW signal, also known as the GW memory effect, is visible as an offset in the GW waveform during the explosion phase, as we can see for, e.g., $12M_\odot$ model. Depending on the model, the offset can be as high as $hD\sim 3 \,\mathrm{cm}$. In contrast to this model, the shock in the $9M_\odot$ model remains quasi-spherical during the early explosion phase. As a result, this model does not exhibit any offset in the GW strain. In addition to the offset, the rest of the models exhibit high-frequency variations even after transitioning to explosion. This behavior is caused by the aspherical expansion of the shock, which produce funnels of accretion onto the PNS, triggering its oscillations and producing GWs. 

\section{Rapidly Rotating Case}
\label{sec:rotating_case}

While most stars are expected to rotate slowly, in order to explain the energetics of the most powerful ${\sim} 10^{52}\,\mathrm{erg}$ explosions, some stars must be rapid rotators. Below, we discuss the GW signature of such stars. 

\subsection{Bounce and ring-down signal}
\label{sec:hydro.bounce}

\begin{figure}
\begin{center}
 \includegraphics[angle=0,width=0.49\columnwidth,clip=false]{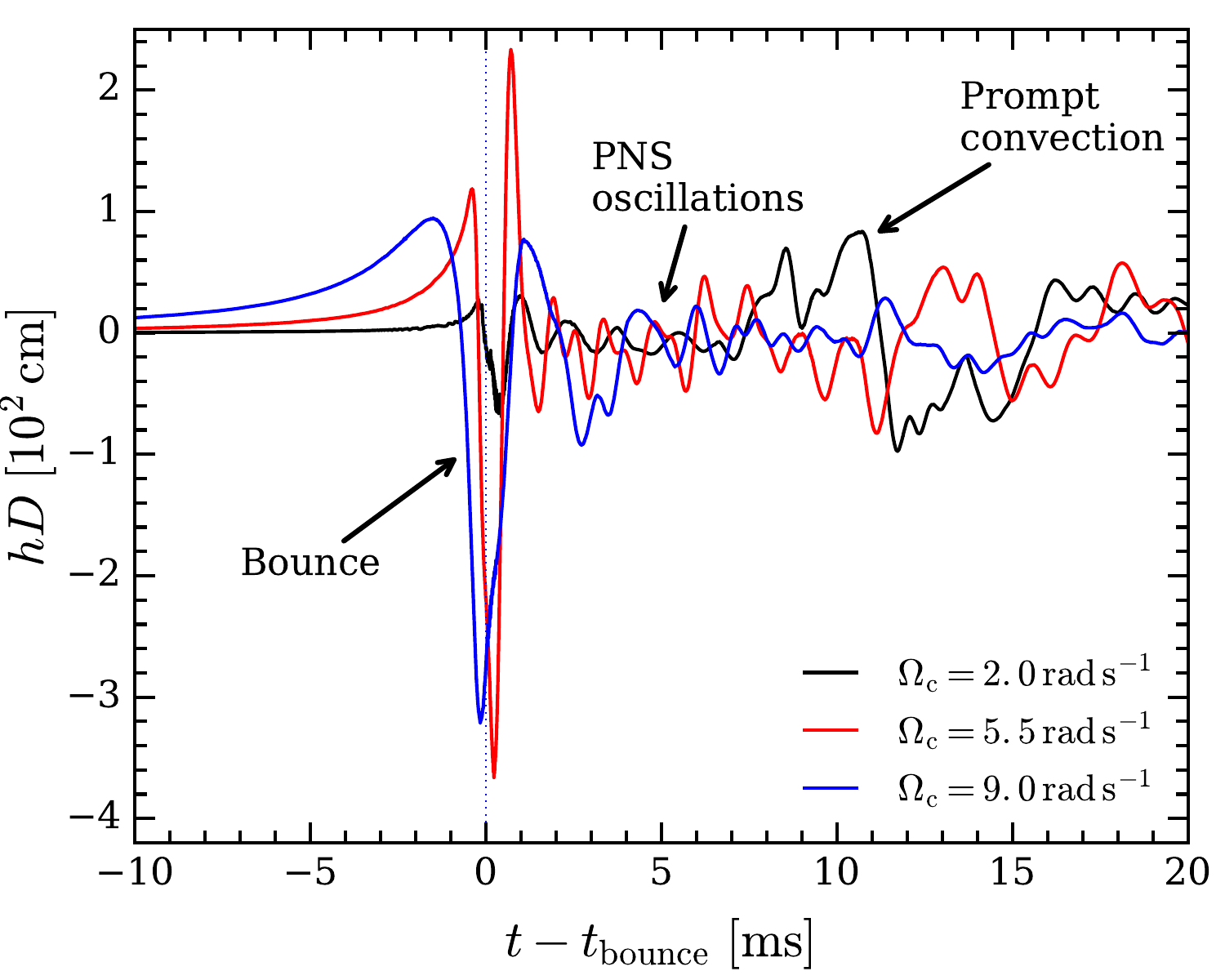}
 \includegraphics[angle=0,width=0.49\columnwidth,clip=false]{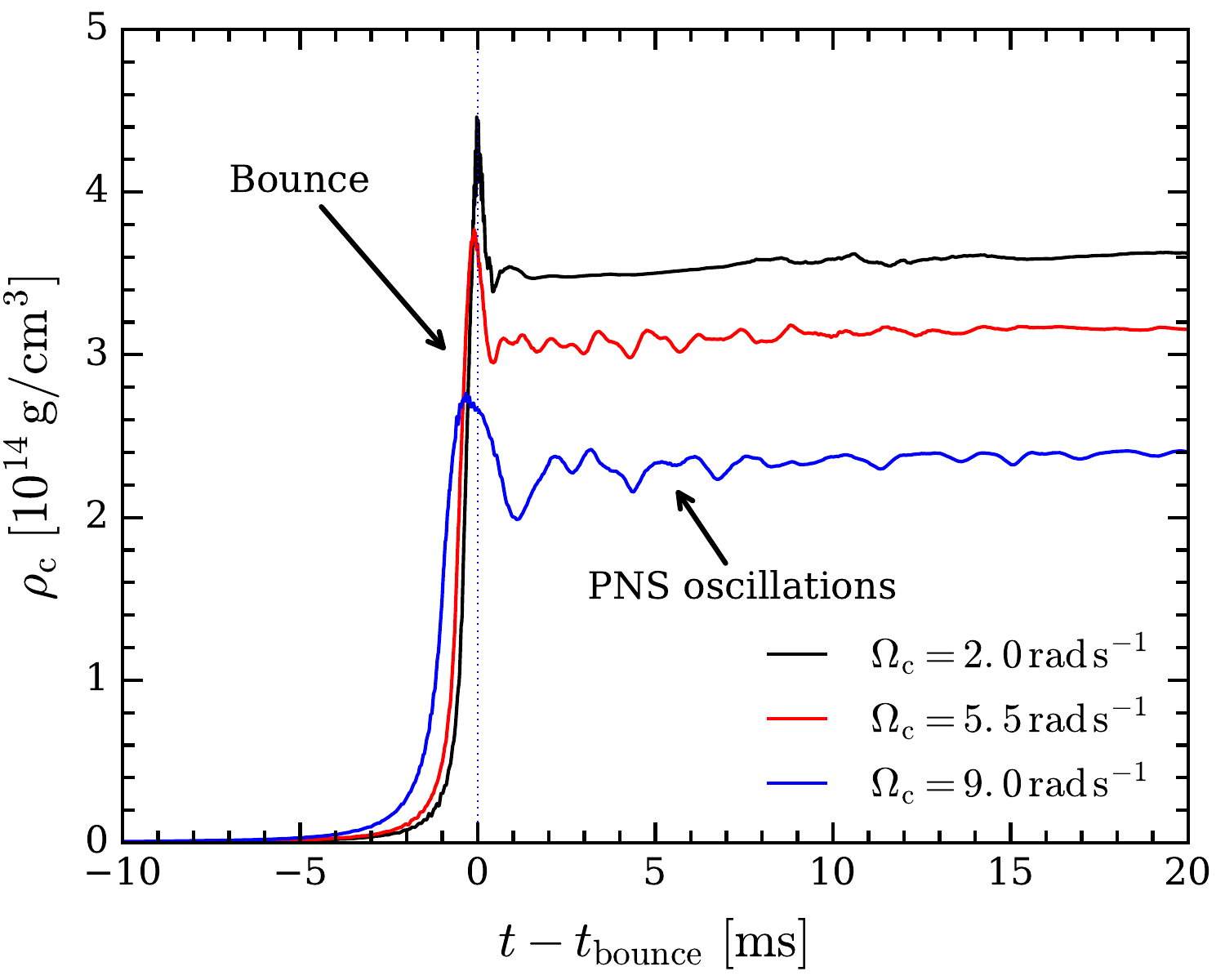}
  \caption{GW strain ({\bf left panel}) and central density  ({\bf right panel}) as a function of time during late collapse, bounce, and early post-bounce phase for three models with slow, rapid, and extremely rapid rotation. The data is produced by \cite{abdikamalov:14}.  
  \label{fig:gw_vs_t}}
\end{center}
\end{figure}

Due to centrifugal support, the core collapse is slower along the equatorial plane than along the rotation axis. As a result, the PNS is born with axisymmetric oblate $\ell=2$ deformation at bounce \cite{ott:12}. The ensuing PNS oscillations last for ${\sim} 10{-}20\,\mathrm{ms}$. Due to the axisymmetric geometry of the initial perturbation, the PNS pulsations remain axisymmetric in this phase (hence, 2D axisymmetric simulations can accurately predict the signal in this phase). In contrast to the perturbations in slowly or non-rotating models, the centrifugal deformation is not stochastic. As a result, the bounce GW signal in rapidly rotating case is ``deterministic'' for a given set of model parameters (and for a given model of nuclear and neutrino physics). 

The degree of centrifugal deformation, and thus the strength of the GW signal, grows with rotation. However, at extremely rapid rotation, the centrifugal support slows the collapse and prevents PNS from reaching higher densities. This limits the GW amplitudes. We can see this trend in Fig.~\ref{fig:gw_vs_t}, which shows the GW strain along equatorial plane (left panel) as a function of time for three different models with pre-collapse central angular velocity of $2.0$, $5.5$, and $9.0\,\mathrm{rad/s}$ for a $12M_\odot$ progenitor \cite{abdikamalov:14}. The peak GW frequencies of these models are about $800$, $750$, and $372$ Hz \cite{abdikamalov:14}. In these models, the rotation respectively has little, strong, and dominant impact of the dynamics of the system. We refer to these models as slowly, rapidly, and extremely rapidly rotating models, respectively. As expected, the slowest rotating model produces little GW signal at bounce. However, it develops  prompt convection within ${\sim} 10\,\mathrm{ms}$ of bounce, which can be seen in the GW signature in Fig.~\ref{fig:gw_vs_t}. The rapidly rotating models do not exhibit prompt convection because these models have strong positive gradient of specific angular momentum in the post-shock region: a fluid element from an inner region cannot easily rise due to their smaller centrifugal support. This hinders convection. In the model with extreme rotation, we can see that the bounce spike is wider, implying slower dynamics and lower frequencies. Due to the axisymmetric geometry of the bounce and ring-down pulsations, the GW signals are only visible for observers close to the equatorial plane and no signal is emitted along the rotation axis \cite{ott:09review}. 

The evolution of the central density, shown on the right panel of Fig.~\ref{fig:gw_vs_t}, reveals that the central density is lower in models with faster rotation. This is expected as stronger centrifugal support makes the PNS less compact. We can also see that the slowly rotating models do not exhibit strong oscillations in $\rho_\mathrm{c}$ in the post-bounce phase. That is because, due to small centrifugal support, this model is born with little oblate deformation. Hence, the PNS settles to a quasi-equilibrium configuration within ${\sim}1\,\mathrm{ms}$ after bounce. In rapidly rotating models, $\rho_\mathrm{c}$ exhibits strong oscillations as the PNS is born with large centrifugally induced oblate deformations. 

We can get a more detailed understanding of the impact of rotation if we look at a large sequence of models with varying degrees of rotation and its distribution. This has been attempted by \cite{abdikamalov:14}, who studied a set of 98 rotational configurations of the $12M_\odot$ model using the two-parameter law for angular velocity:
\begin{equation}
    \Omega(\varpi) = \Omega_\mathrm{c} \left[1+\left(\frac{\varpi}{A}\right)^2\right]^{-1}
    \label{eq:rotation_law}
\end{equation}
where $\Omega_\mathrm{c}$ is the pre-collapse central angular velocity, $\varpi$ is the distance from the rotation axis, while $A$ is a characteristic distance from the rotation axis over which the angular velocity decreases by a factor of $2$. They considered five different values of $A$ ranging from $300\,\mathrm{km}$, which represents extremely strong differential rotation, to $10,000\,\mathrm{km}$, which represents  weak differential rotation. 

\begin{figure}
\begin{center}
 \includegraphics[angle=0,width=0.8\columnwidth,clip=false]{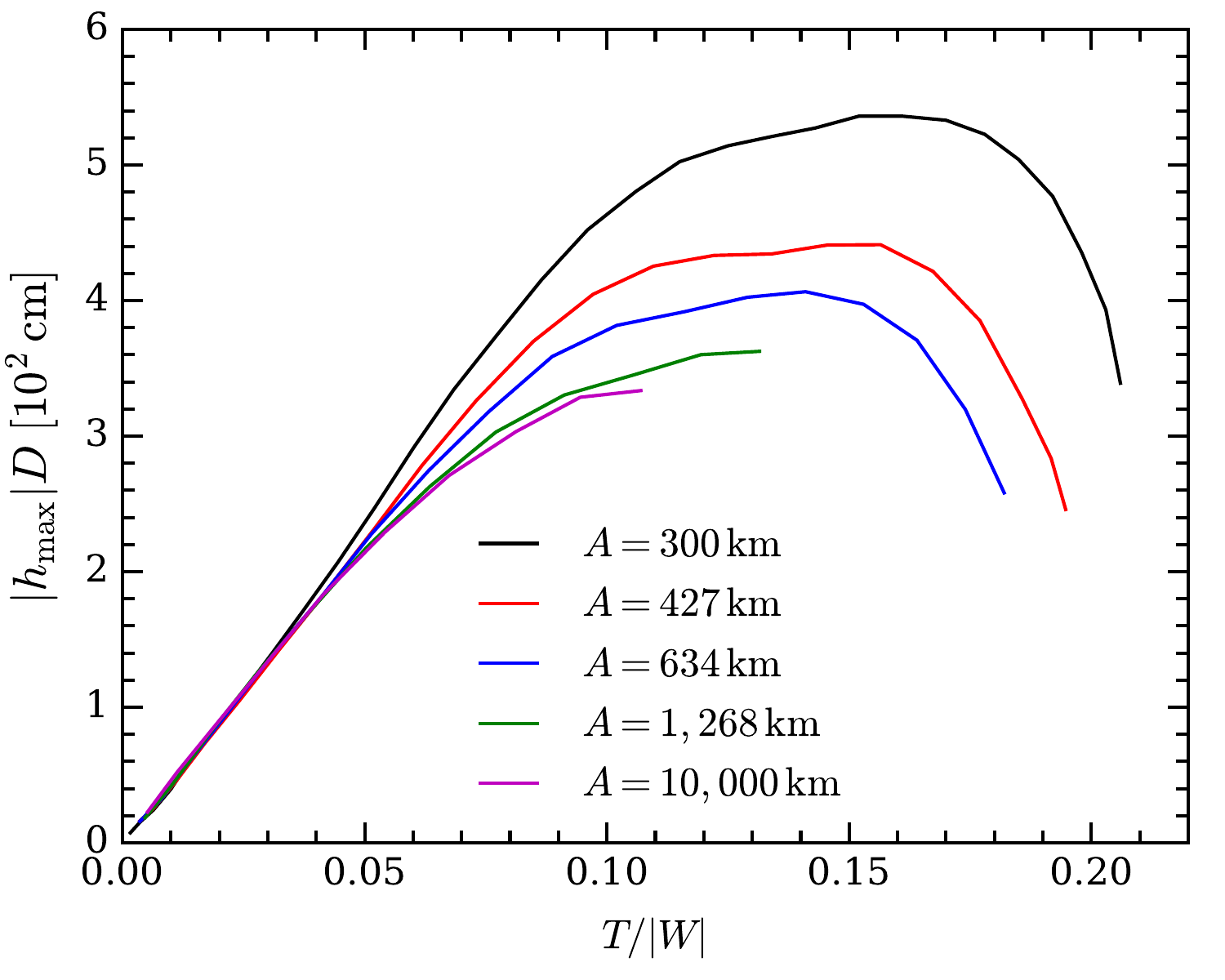}
  \caption{The maximum GW strain produced by core bounce as a function of the rotation parameter $T/|W|$ of the inner core at bounce for models with different degrees of rotation. Here, $A$ is the degree of differential rotation (cf. eq.~\ref{eq:rotation_law}). The GW waveforms are produced by \cite{abdikamalov:14}.  
  \label{fig:hmax_vs_betaic}}
\end{center}
\end{figure}

Figure~\ref{fig:hmax_vs_betaic} shows the maximum GW strain produced at bounce as a function of the ratio $T/|W|$ of the rotational kinetic energy $T$ and potential binding energy $|W|$ of the inner core at bounce. Because this parameter is a ratio of the two energies, it characterizes the dynamical importance of rotation. As we can see, the GW strain $h$ increases linearly with $T/|W|$ for $T/|W| \lesssim 0.06{-}0.09$, but the growth slows for larger $T/|W|$, and then $h$ starts even decreasing with $T/|W|$ for $T/|W|\gtrsim 0.17$. Also, for $T/|W| \lesssim 0.06{-}0.09$, the GW signal is not sensitive to the degree of differential rotation $A$, while for higher $T/|W|$, the GW signal depends on $A$. 

The linear increase of $h$ with $T/|W|$ for $T/|W| \lesssim 0.06{-}0.09$ is easy to understand. Since $|W| \sim GM^2/R$ and $E_\mathrm{kin} \sim T$, the estimate (\ref{eq:h_est}) leads to 
\begin{equation}
    hD \sim \frac{GM \Omega^2 R^2}{c^4} \sim \frac{(GM)^2}{c^4 R} \frac{T}{|W|},\label{eq:gw_vs_tw}
\end{equation}
Thus, the GW signature of the bounce and ring-down oscillations is mostly determined by the mass, radius, and the rotation parameter of the inner core. The fact that, for $T/|W|\lesssim 0.06{-}0.09$, the GW strain is so well approximated by simple relation (\ref{eq:gw_vs_tw}) suggests that the dynamics is governed by the fundamental quadrupole mode oscillations of the PNS. This is supported by detailed oscillation mode analysis \cite{fuller:15, richers:17}. The mode frequency is comparable to the dynamic frequency of the PNS, 
\begin{equation}
f_\mathrm{peak} \sim \frac{1}{2\pi} \sqrt{G\rho_\mathrm{c}},
\end{equation}
where $\rho_\mathrm{c}$ is the PNS central density \cite{richers:17}. For a typical PNS, $f_\mathrm{peak} \sim 750$ Hz.

For more rapid rotation, $T/|W| \gtrsim 0.06{-}0.09$, the $h$ \emph{vs.} $T/|W|$ curve in Fig~\ref{fig:hmax_vs_betaic} deviates from a straight line. The peak frequency of the signal $f_\mathrm{peak}$ becomes larger than the dynamical frequency $\sqrt{G\rho_\mathrm{c}}/2\pi$. More specifically, $f_\mathrm{peak}$ now scales as 
\begin{equation}
    f_\mathrm{peak} \sim \frac{1}{2\pi} \frac{\sqrt{G \rho_\mathrm{c}}+\Omega_\mathrm{max}}{2},
\end{equation}
for $0.06{-}0.09 \lesssim T/|W| \lesssim 0.17$, where $\Omega_\mathrm{max}$ is the highest angular velocity achieved outside of radius of $5\,\mathrm{km}$ \cite{richers:17}. The transition happens when $\Omega_\mathrm{max}$ becomes larger than $\sqrt{G \rho_\mathrm{c}}$, which is a regime where Coriolis force becomes dynamically important. These observations suggest that the dynamics of the PNS is not dominated by the fundamental quadrupole mode anymore. Instead, the PNS oscillations may be transiting to a mode (or modes) supported by the Coriolis force. At extremely rapid rotation of $T/|W| \gtrsim 0.17$, the dynamics is dominated by the centrifugal force. This leads to complex PNS behavior involving multiple modes with comparable amplitudes. The exact nature of the PNS oscillation modes in this regime is yet to be established.

Despite its short ${\sim}20\,\mathrm{ms}$ duration, the GW signal from rotational bounce and post-bounce ring-down oscillations is detectable up to distance of $\sim 50\,\mathrm{kpc}$ with current detectors \cite{gossan:16}. Once detected, for rapidly rotating models with optimum orientation, it is possible to extract parameters of the CCSN central engine, such as rotation \cite{abdikamalov:14}. A possibility of extracting the parameters of high-density nuclear matter has also been explored \cite{richers:17}.

\subsection{Non-axisymmetric instabilities}

Sufficiently rapidly rotating stars may be subject to non-axisymmetric instabilities. These instabilities may last for many dynamical timescales, leading to long-lasting GW emission. There are different types of non-axisymmetric instabilities\footnote{We focus only on the instabilities that may occur in the context of CCSNe. Outside this context, many other types of instabilities have been discussed, especially in the context of accretion disks.}. Based on their growth rate, instabilities can be divided into two sub-groups: secular and dynamical. 

The secular instability is driven by dissipation (viscosity or GW emission) and develops on a dissipation timescale \cite{shapteu:83}, which is ${\sim} 1\,\mathrm{s}$ or longer for CCSNe. The viscosity drives a non-axisymmetric instability for the $\ell=-m$ modes if their frequencies pass through zero in the corotating frame \cite{paschalidis:17}. For modes that are counter-rotating in the frame of the star, but appear corotating to a faraway inertial observer, the GWs drive an instability via the Chandrasekhar-Friedman-Schutz  mechanism. GWs extract the stellar angular momentum by making the mode angular momentum increasingly negative. The $f$, $r$, and $w$ modes have been discussed in this context (see, e.g., \cite{paschalidis:17} for a recent review). Within ${\sim} 1\,\mathrm{s}$ after formation, PNSs undergo changes that are faster than the secular timescale (e.g., accretion of mass and angular momentum), which may interfere with the development of the instability. Also, current CCSN simulations do not cover timescales beyond ${\sim} 1\,\mathrm{s}$. For this reason, such instabilities have yet not been observed. Nevertheless, one cannot exclude that these instabilities may develop in a later phase, when the PNS evolves at a slower rate.  

The dynamical instability develops on the dynamical timescale $(G\rho)^{-1/2}\sim 1\,\mathrm{ms}$. Two sub-types have been discussed in the context of CCSNe. These are often called high- and low-$T/|W|$ instabilities. The former, also known as dynamical bar mode instability, develops in stars when $T/|W|$ becomes larger than $\sim 0.24{-}0.27$ depending on the compactness of the star \cite{shibata:00}. The low-$T/|W|$ instability develops at lower $T/|W|$, but requires strong differential rotation \cite{centrella:01}. A recent comprehensive review of the instabilities can be found in \cite{paschalidis:17}.

Using the Newtonian quadrupole formula (\ref{eq:h_est}), we can obtain an estimate of the GW strain \cite{ott:09review}: 
\begin{equation}
    h \sim 6 \times 10^{-21} \left(\frac{\epsilon}{0.1}\right)
    \left(\frac{f}{500\,\mathrm{Hz}}\right)^2
    \left(\frac{D}{10\,\mathrm{kpc}}\right)^{-1}
    \left(\frac{M}{M_\odot}\right)
    \left(\frac{R}{12\,\mathrm{km}}\right)^2,
\end{equation}
where $\epsilon$ is the ellipticity of the bar. The luminosity of GWs is
\begin{equation}
    \frac{dE_\mathrm{GW}}{dt}\simeq \frac{32}{5} \frac{G}{c^5} I^2 \epsilon^2 \Omega^6,
\end{equation}
where $I$ is the moment of inertia of the star with respect to the rotation axis \cite{shapteu:83}. If the deformation persists for $N$ rotation periods, then the SNR of the signal increases by a factor of $N^{1/2}$, significantly improving the detectability \cite{fryer:02, kotake:17}. 

The highest $T/|W|$ observed in CCSN simulations barely exceeds $0.22$ \cite{dimmelmeier:08}. During collapse, due to conservation of angular momentum, the angular velocity increases as $r^{-2}$. This means that centrifugal force is $\propto r^{-3}$. On the other hand, the gravity is $\propto r^{-2}$. Therefore, for sufficiently rapid rotation, the centrifugal support becomes stronger than gravity, limiting collapse. This barrier limits the value of $T/|W|$ to $\simeq 0.22$, which is below the threshold for the dynamical high-$T/|W|$ instability. As a result, the high-$T/|W|$ dynamical instability is unlikely to develop in CCSNe \cite{kotake:13review}. 

The low-$T/|W|$ instability develops in the presence of a corotation radius, a radius where rotation matches the speed of the mode \cite{watts:05}. This is possible only in stars with differential rotation. The low-$T/|W|$ instability has been observed in a number of simulations (e.g., \cite{shibagaki:20} and references therein). The $m=1$ or $m=2$ modes are often dominant, but higher-$m$ modes were also observed, albeit with smaller amplitudes \cite{kotake:13review}. 

\begin{figure}
\begin{center}
 \includegraphics[angle=0,width=0.9\columnwidth,clip=false]{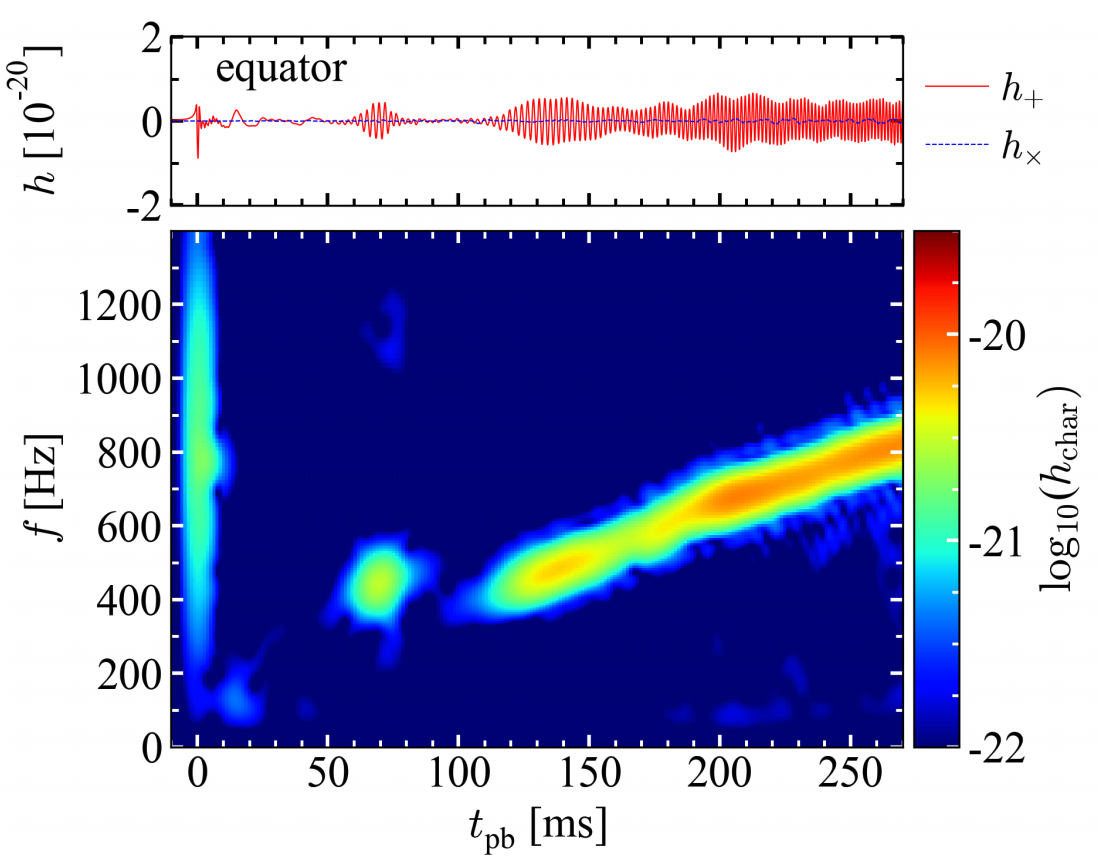}
  \caption{GW strain amplitude ({\bf top panel}) and GW strain spectrogram ({\bf bottom panel}) along the equatorial plane as a function of time after bounce. On the left panel, we can easily see $m=1$ deformation, while on the right panel, we can identify $m=2$ deformation. Adapted from Shibagaki, Kuroda, Kotake, Takiwaki, \textit{A new gravitational-wave signature of low-$T/|W|$ instability in rapidly rotating stellar core collapse}, MNRAS \textbf{493} 138 (2020) \cite{shibagaki:20}. \textcopyright~Shibagaki et al. Reproduced with permission.
  \label{fig:toverw_gw}}
\end{center}
\end{figure}

\begin{figure}
\begin{center}
 \includegraphics[angle=0,width=1.0\columnwidth,clip=false]{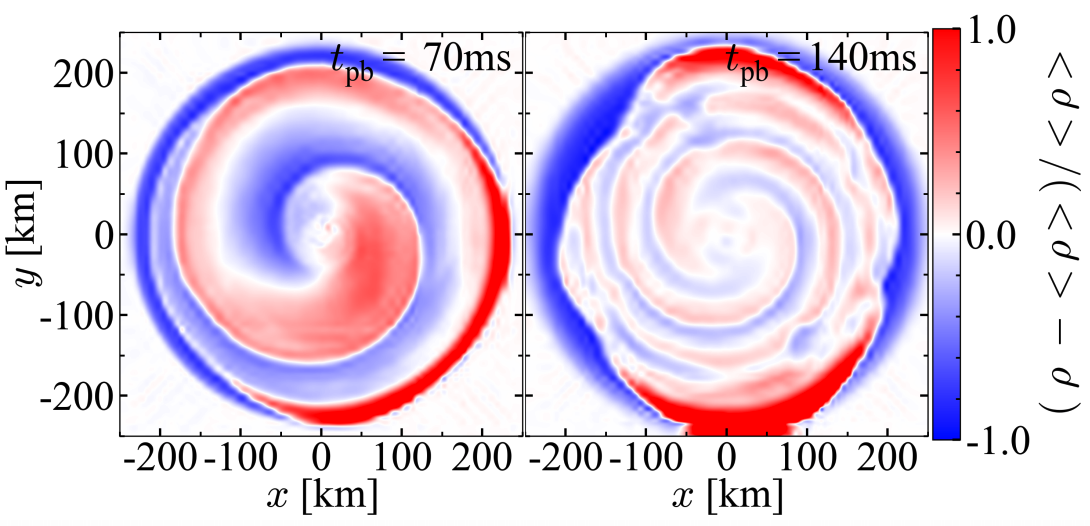}
  \caption{Deviations of density from axisymmetric average at $70\,\mathrm{ms}$ ({\bf left panel}) and $140\,\mathrm{ms}$ ({\bf right panel}) after bounce. The emission at $\simeq 60-70\,\mathrm{ms}$ after bounce is caused by the $m=1$ mode, while the signal after $\simeq 120\,\mathrm{ms}$ is caused by the $m=2$ mode. Adapted from Shibagaki, Kuroda, Kotake, Takiwaki, \textit{A new gravitational-wave signature of low-$T/|W|$ instability in rapidly rotating stellar core collapse}, MNRAS \textbf{493} 138 (2020) \cite{shibagaki:20}. \textcopyright~Shibagaki et al. Reproduced with permission.  
  \label{fig:toverw_rho}}
\end{center}
\end{figure}

As an example, below we discuss a model with initial mass of $70M_\odot$ that has been simulated by \cite{shibagaki:20} until $270\,\mathrm{ms}$ after bounce. Before collapse, this model has a central angular velocity of $2\,\mathrm{rad/s}$ and $T/|W|$ of $\sim 3\times 3^{-3}$. In the post-bounce, $T/|W|$ becomes $\sim 0.05$. The top and bottom panels of Fig.~\ref{fig:toverw_gw} show the GW strain and spectrograms observed along equatorial plane as a function of time. The initial rotational bounce and ring-down signal within ${\sim} 20\,\mathrm{ms}$ after bounce is followed by a quiescent period that lasts until $60\,\mathrm{ms}$ after bounce. Between $60$ and $80\,\mathrm{ms}$, we see GW emission with frequency ${\sim} 400\,\mathrm{Hz}$. The density deviations from axisymmetry, shown in Fig~\ref{fig:toverw_rho} at $t_\mathrm{pb} = 70\,\mathrm{ms}$ at the equatorial plane, reveals $m=1$ deformation. Another quiescent period is observed between $80$ and $140 \,\mathrm{ms}$, after which another mode develops and persists until the end of simulation at $t_\mathrm{pb}=240\,\mathrm{ms}$. The density deviations at $t_\mathrm{pb}=140\,\mathrm{ms}$ reveal $m=2$ deformation. In this phase, the GW frequency increases from ${\sim} 400$ until ${\sim} 800\,\mathrm{Hz}$ by the end of the simulation. This is caused by cooling of the star, which leads to contraction and increase of the pattern speed of the mode. 

This resulting signal is easily detectable anywhere within our galaxy with current detectors. The future third generation may be able to detect up to $\sim 1\, \mathrm{Mpc}$ distances \cite{shibagaki:20}. The event rate is not precisely known, but we can make rough estimate. In the local group, the CCSN rate is ${\sim} 20{-}80$ events per $100$ years \cite{Nakamura:2016kkl}. Assuming that $\sim1\%$ of all CCSNe rotate rapidly enough to produce low-$T/|W|$ instability, we arrive at ${\sim} 0.2{-}0.8$ events per $100$ years.

Despite significant progress in our understanding of non-axisymmetric instabilities, there are aspects that could benefit from further exploration. In particular, more realistic simulations for a large set of progenitors will shed light on the precise conditions for the development and growth of the instability as well as on the saturation amplitude and the persistence of the non-axisymmetric modes. Here, it is crucial to accurately capture the accretion of angular momentum onto PNS as well as the transfer of angular momentum from the PNS to outer regions by neutrinos and magnetic fields. This is a subject of ongoing research.

\subsection{Collapse to black hole}
\label{sec:bh_collapse}

When a PNS collapses to a BH, the collapsing star emits a GW spike at its dynamical frequency of ${\sim} 1\,\mathrm{kHz}$, followed by quasi-normal modes of the newly formed BH at a few kHz \cite{cerda:13}. Due to the short duration and the high frequency of the signal, the BH formation signal will be challenging to detect.

\section{Anisotropic Neutrino Emission}
\label{sec:numem}

Anisotropic emission of neutrinos produces a flux of outgoing energy with non-uniform angular distribution, which represents a time-changing quadrupole moment. This produces a slowly varying non-oscillatory contribution to the GW strain, somewhat similarly to the way aspherical shock expansion produced an offset in the GW strain. If $E_\nu$ is the total energy of emitted neutrinos and $\Delta E_\mathrm{ani} \sim \alpha_\nu E_\nu$ is energy variation due to anisotropy, the resulting signal GW strain can be estimated to an order of magnitude using (\ref{eq:h_est}):
\begin{equation}
   hD \sim \frac{2G}{c^4} \alpha_\nu E_\nu
\end{equation}
In the context of CCSNe, this estimate can be rewritten as \cite{mueller:97}
\begin{equation}
    h D \sim 1.6\times10^2 
    \left(\frac{\alpha_\nu}{10^{-2}}\right) \left(\frac{L_\nu}{10^{53}\,\mathrm{erg/s}}\right)
    \left(\frac{\Delta t}{1\,\mathrm{s}}\right) \,\mathrm{cm},
\end{equation}
Using similar argument, we can obtain an estimate for GW energy
\begin{equation}
    E_\mathrm{GW,\nu} \sim 10^{-8} \alpha_\nu^2 
    \left(\frac{L_\nu}{10^{53}\,\mathrm{erg/s}}\right)^2
    \left(\frac{\Delta t}{1\,\mathrm{s}}\right)
    M_\odot c^2.
\end{equation}
where $L_\nu$ is the total neutrino luminosity. The anisotropy in neutrino emission can be produced by deformations induced by rotation and/or multi-dimensional hydrodynamic instabilities. For $L_\nu \sim 10^{53}\,\mathrm{erg/s}$, which is typical during the first second after bounce, and for $\alpha_\nu\sim 10^{-2}$, we find $hD\sim 10^2\,\mathrm{cm}$, which is in line with more detailed calculations  \cite{mueller:97}. This is comparable to the GW strain produced by asymmetric matter motions in CCSNe. However, the quadrupole moment of a system of anisotropically propagating neutrinos changes slowly with time. As a result, the energy of GWs is ${\sim} 10^{-12} M_\odot c^2\simeq 1.8\times10^{42}\,\mathrm{erg}$, which is ${\sim} 4$ orders of magnitude smaller than that from the matter motion. For this reason, and due to the non-oscillatory nature of the signal, the GW signal from the anisotropic neutrino emission is unlikely to be detectable with current detectors. However, detection could be possible with the future third-generation detectors. In order to make a more definitive statement, the CCSN simulations need to cover timescales longer than ${\sim} 1\,\mathrm{s}$, which will reveal more precise time evolution of the GW signal from the anisotropic neutrino emission. 

\section{Quark deconfinement phase transition}
\label{sec:phase_transition}

When the central density of the PNS is sufficiently large, the nuclear matter may undergo quark-deconfinement phase transition. The exact details of whether and how this happens are not well established. If the phase transition leads to rapid and strong pressure reduction, it may cause a ``mini-collapse'' of a PNS to a more compact configuration. In rapidly rotating models, the mini collapse may excite the fundamental quasi-radial and and quadrupole oscillation modes of the PNS, which leads to periodic signal with a ${\sim} 1\,\mathrm{kHz}$ frequency \cite{abdikamalov:09}. In slowly rotating models, the radial mini-collapse may interact with the asphericities of the flow outside of PNS and lead to strong GW emission \cite{zha:20}. For a sufficiently strong mini-collapse, such a signal could be detectable for a source within our galaxy. 

\section{Multi-Messenger Aspects}
\label{sec:mma}
CCSNe represent a perfect target for a multi-messengers study. 
Indeed, the enormous energy developed by the stellar explosion, $\sim 10^{53}$ erg, is released through different channels. 
About $99\%$ of the energy is converted in low-energy (MeV) neutrinos, the leftover ${\sim} 1\%$ is mainly kinetic energy of the shock wave, while approximately the $0.01\%$ of this energy powers a multi-wavelength electromagnetic (EM) emission and only a few times $10^{46}$ erg is released through gravitational waves as discussed in previous sections. 
\begin{figure}[h]
    \centering
    \includegraphics[width=\textwidth]{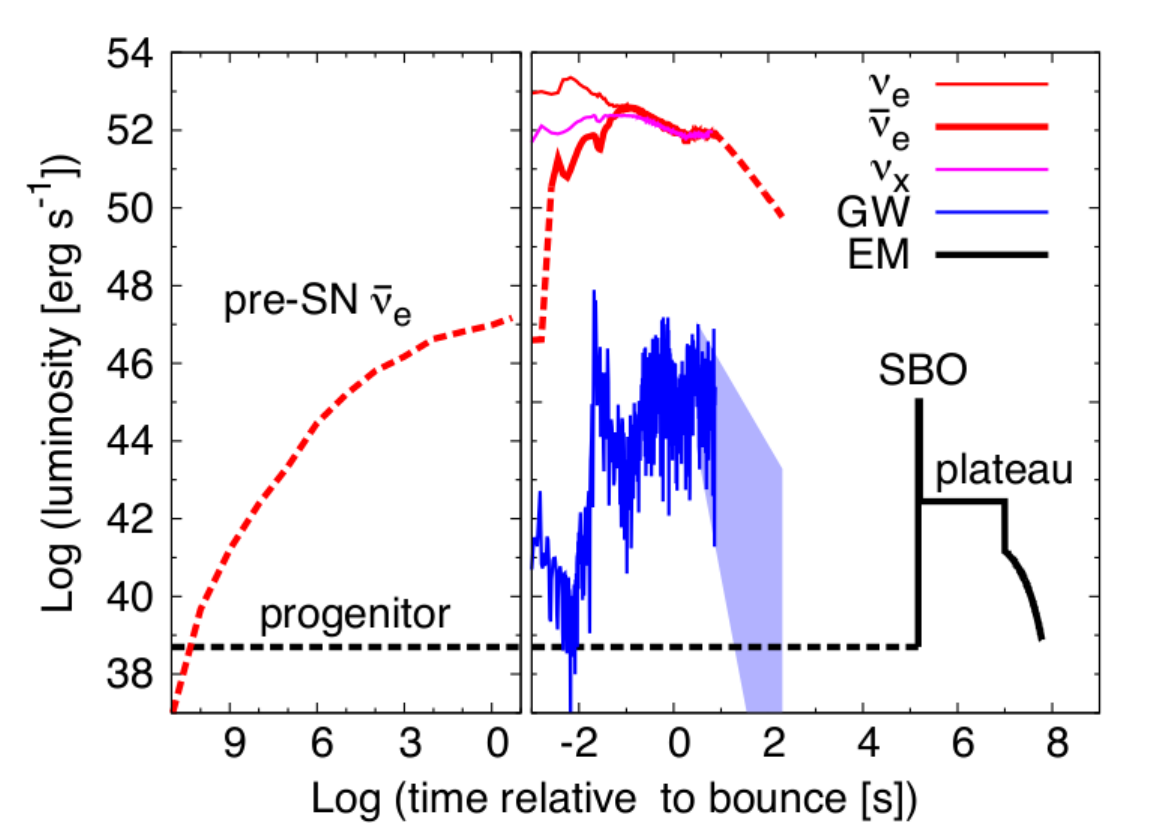}
    \caption{Time evolution of neutrino, GW (thin blue line), and EM (thick black line) signal luminosities for a non-rotating $17 M_\odot$ progenitor of \cite{Nakamura:2016kkl}. Thin and thick red lines represent $\nu_e$ and $\bar{\nu}_e$, while magenta line is for $\nu_x$, which represents $\nu_\mu$, $\nu_\tau$, $\bar{\nu}_\mu$, or $\bar{\nu}_\tau$. The solid lines are from simulations, while dashed lines are based on approximate estimates. The left and right panels represent time before and after core bounce, respectively. 
    Reproduction of Fig.1 from K.~Nakamura, S.~Horiuchi, M.~Tanaka, K.~Hayama, T.~Takiwaki and K.~Kotake, \textit{Multimessenger signals of long-term core-collapse supernova simulations: synergetic observation strategies}, MNRAS \textbf{461} 3 (2016)\cite{Nakamura:2016kkl} \textcopyright~Nakamura et al. Reproduced with permission.}
    \label{fig:multimessenger}
\end{figure}
The time evolution of the multi-messenger signals expected from a CCSN explosion is reported in Fig.~\ref{fig:multimessenger}. These signals are obtained from the numerical simulation of a neutrino-driven explosion of a non-rotating $17 M_\odot$ progenitor \cite{Nakamura:2016kkl}.

The EM burst of a supernova starts at shock breakout (SBO), when the shock emerges from the stellar surface hours after core collapse (cf. Fig.~\ref{fig:multimessenger}). Characterized by a flash of UV and X-rays, it could provide important information about the CCSN progenitors, such as radius \cite{Matzner:1998mg}. However, SBO detection is challenging since it is a short-lived phenomenon that can last only hours to days, depending on the density at shock emergence \cite{Waxman:2016qyw}. After the SBO, the EM emission enters the plateau phase lasting about 100 days. During this phase, the luminosity and duration of the plateau can also provide interesting constraints on the progenitor, like its radius and the ejecta mass \cite{Nakamura:2016kkl}.   

As a last remark, it is worth noting that EM emission could be absent (or too faint to be detected). In a small fraction of CCSNe, known as ''failed'' supernovae, the massive star experiences core collapse and bounce, nevertheless is unable to generate a successful explosion and ends up with a BH formation. If EM signal is too faint, GWs and/or neutrinos detection represents the only possibility to directly investigate these astrophysical events.

As stated above, neutrinos and antineutrinos of all flavors are copiously produced and released during a CCSN event. 
They are characterized by energies in the few to tens of MeV range. The total neutrino burst duration is a few tens of seconds.
Neutrino luminosities are drawn with red lines in Fig.~\ref{fig:multimessenger} and the reference time is the time of the bounce. 
The solid lines represent results from numerical simulation, whereas the dashed lines are approximate estimates \cite{Nakamura:2016kkl}. 
The left-hand panel of Fig.~\ref{fig:multimessenger} shows the pre-SN emission of $\bar{\nu}_e$ due to the late silicon burning phase, 
that can release $\sim 5\times 10^{52}$ erg in neutrinos during the few hours just before the star collapse. The detection of this neutrino signal could be exploited as a very useful early warning alert \cite{Odrzywolek:2003vn}. 
In the right-hand panel of Fig.~\ref{fig:multimessenger} neutrino emission can be described as composed of three different temporal phases: neutronization, accretion and cooling \cite{janka:12a}. 
In the shocked material, electron capture by protons produces a huge amount of electron neutrinos via reaction $e^{-} +p \rightarrow n+\nu_e$.
Furthermore, the concentration of the positrons, produced through $\gamma\rightarrow e^{+}+e^{-}$, progressively increases. 
These positrons interact with electrons, producing neutrinos of all flavors through the reaction $e^{+} e^{-}\rightarrow \nu_{i}\bar{\nu_i}$ with $i = e;\mu;\tau$. At high material density, neutrinos are trapped. 
During the early shock expansion phase, the density at the shock decreases approaching the value $\rho \simeq 10^{12} \mathrm{g/cm^3}$ at which point neutrinos are free to escape. This prompt neutrino emission, called neutronization burst, is characterized by a sharp peak in the electron neutrino luminosity lasting a few milliseconds and releasing an energy of the order $10^{51}$~erg. 
As described in the previous sections, the shock wave stagnates and transforms into an accretion shock. In the nearly transparent shocked region, around the high-density core of the PNS, $\bar{\nu_e}$ and $\nu_e$ are produced by the symmetric processes  $e^{-} +p \rightarrow n+ \nu_e$ and  $e^{+} +n \rightarrow p+ \bar{\nu_e}$. 
This emission generates a prominent hump in the neutrino and antineutrino electronic flavour luminosities lasting a fraction of a second before the final star explosion. Finally the PNS evolves to a hot neutron star that cools down by emitting neutrinos and antineutrinos of all species produced by neutral current processes. During this Kelvin-Helmholtz cooling phase, neutrinos carry away $90\%$ of the total energy emitted in the CCSN with a characteristic diffusion time of a few seconds.

Neutrinos and antineutrinos, produced inside the CCSN by all the interaction processes described above, undergo flavour conversion while propagating outward through the star and interstellar medium. The oscillation probability depends on the neutrino mass hierarchy and can be affected by complex non-linear terms due to neutrino-neutrino interactions. 
In a simple scenario, the Mikheyev-Smirnov-Wolfestein matter effect will mix flavours during propagation in such a way that the electronic antineutrino flux arriving to the Earth is a mixture of the electronic antineutrinos and heavy lepton flavour fluxes at the production, i.e. $\Phi_{\bar{\nu_e}}=\bar{P} \Phi^0_{\bar{\nu_e}} + (1-\bar{P}) \Phi^0_{\bar{\nu_x}}$, where $\bar{P}$ is the survival probability of $\bar{\nu_e}$. A similar expression holds for electronic neutrinos, i.e. $\Phi_{\nu_e}=P \Phi^0_{\nu_e} + (1-P) \Phi^0_{\nu_x}$, where $P$ is the survival probability of $\nu_e$. Approximated numerical values for the survival probabilities are $\bar{P}\simeq 0.7(\simeq 0)$ and $P\simeq 0 (\simeq 0.3)$ for the normal (inverted) mass hierarchy. This implies that the flux of electronic neutrinos will suffer a strong suppression during propagation. In particular, emission phases of pure $\nu_e$ (e.g. the neutronization burst) can mainly be investigated by the detection of the $\nu_x$ flux.   
 
\subsubsection{GW Searches}
The GW signals expected from CCSNe have been deeply discussed in the previous sections of this chapter. 
They can originate from a variety of mechanisms and can be characterized by very different frequency ranges that may or may not fall within the sensitive frequency bands of GW interferometers. Moreover, GWs from CCSNe have the common feature of being "short-burst like", i.e., impulsive signals lasting less than a second. This means that they can be easily confused with the non-stationary part of the noise of GW interferometers, known as glitches. 
In order to maximize the ability to detect this kind of GWs, the Advanced LIGO and Advanced Virgo detectors \cite{TheLIGOScientific:2014jea,TheVirgo:2014hva} perform all-sky searches of short-duration signals with no prior assumption on the GW signal time of arrival or sky direction \cite{Abbott:2019prv}. One of the three pipelines used to search short-burst is the Coherent WaveBurst (cWB) pipeline \cite{Klimenko:2015ypf}, specifically designed to perform searches for unmodelled bursts with a duration of up to a few seconds in the frequency range $32{-}4096$ Hz. In this pipeline, coincident events are ranked according to their coherent network signal-to-noise ratio (SNR). This coherent SNR should be higher for cross-correlated signals and lower for uncorrelated glitches, thus favoring real GW signals. The background distribution of triggers is calculated by time-shifting the data of one detector with respect to the other detectors by an amount of time that breaks any correlation between detectors for a real signal.
Real temporal coincidences among interferometers, called 0-lag coincidences, are then compared with the accidental ones and this comparison provides their false alarm rates.

The detection efficiency of all-sky searches, performed by Advanced LIGO and Advanced Virgo, is estimated by injecting simulated GW signals into real detector data and quantifying the percentage of signals recovered by the pipeline with a false alarm rate lower than the threshold of $1/100$ years. Such a sensitivity study is performed by considering a set of ad hoc waveforms, for example Gaussian or Sine-Gaussian signals, that are not derived from specific astrophysical simulations and that can be described by few characteristic parameters, like the central frequency or duration.   
In the most recent published results \cite{Abbott:2019prv}, for a source located at $D=10$ kpc and for a false alarm rate of 1/100 years, $50\%$ detection efficiency is obtained for root-mean-square strain amplitude in the range $(1{-}10)\times 10^{-22}\sqrt{Hz}$ depending on the central signal frequency. This range of values for the root-mean-square strain amplitude can be converted into an equivalent range for the minimum amount of energy emitted by a GW CCSN to be detected. At the present, Advanced LIGO and Advanced Virgo, in the frequency band $(100-200)$ Hz, can reach $50\%$ of detection efficiency for a GW emitted energy $\gtrsim 10^{-9}M_\odot c^2$ and a source emitting at 10 kpc.

Simulated GW waveforms from CCSN, as deeply discussed in the previous sections of this chapter, could be very complex and different depending on the emission mechanisms and the progenitor mass. 
For non-rotating stars, that represent the majority of the cases, the GW emission due to prompt convection and neutrino-driven convection, as the one reported in Fig.~\ref{fig:multimessenger}, 
could be very difficult to identify. Even for a CCSN located in the Galactic Center, i.e., $D=8.5$ kpc, the SNR could be too small to claim a detection \cite{Nakamura:2016kkl}. The capability to separate a real signal from the noise improves when GW search can profit from additional information/constraint provided by the contemporary detection of other probes from the same sources, see, e.g., Fig.~\ref{fig:efficiency}.

\subsubsection{Neutrino Searches}
CCSNe are the only sources of extra-solar low-energy neutrinos detected so far. Indeed, in 1987 the neutrino detectors Kamiokande, IMB and Baksan observed a burst of low-energy neutrinos related to a CCSN explosion in the Large Magellanic Cloud \cite{Pagliaroli:2008ur}. This event, called SN1987A, represents a milestone for the supernova study. It was not only the first CCSN detected by neutrino experiments, but also the first SN visible to the naked eye after the Kepler SN in 1604. The progenitor of SN1987A was found to be a star with mass ${\sim} 15 M_\odot$, located on the outskirts of the Tarantula Nebula in the Large Magellanic Cloud, at a distance of about $50$~kpc. The data collected by the neutrino detectors, despite the small statistics ($\sim 29$ events), confirmed the baseline theory of CCSNe \cite{janka:12a}.

Several of the present neutrino detectors are sensitive to a neutrino burst from a galactic supernova \cite{Scholberg:2012id}. 
The interaction process with the highest cross section in Cherenkov and liquid scintillator detectors is the inverse beta decay (IBD), $\bar{\nu_e}+p\to e^{+}+n$. The expected rate of events could be estimated as $R_{IBD}(t)=N_p\int_{E_{thr}} dE_{\nu}\sigma(E_\nu)\Phi_{\bar{\nu_e}}(E_\nu,t)$
where $N_p$ is the number of proton targets inside the detector, $\sigma$ is the differential cross section of the process and $E_{thr}$ is the energy threshold of the detector. It is important to note that the effects of source distance, $\Phi_{\bar{\nu_e}}\propto D^{-2}$, average energy of neutrino spectrum, $\langle E_{\bar{\nu_e}} \rangle$, and neutrino mass hierarchy are degenerate for this rate.
Let us consider a neutrino detector with $E_{thr}=1$~MeV and hence sensitive to all of the IBD channel spectrum. The total number of events expected for a CCSN located at a distance $D$ is $N_{ev}=\int R_{IBD}(t)dt$ if the average energy of the $\bar{\nu_e}$ at the source is $\langle E_{\bar{\nu_e}} \rangle=12$ MeV and the neutrino mass hierarchy is the inverted one. However, the same number of events $N_{ev}$ is expected also if the source distance is reduced to $0.9 D$ and the neutrino mass hierarchy is the normal one.  
To break this degeneracy more, detectors and/or additional information are needed, such as the "a priori" knowledge of the mass hierarchy or of the CCSN distance, e.g., thanks to the detection of the EM counterpart. 
Finally, several other interaction channels are expected to be observed with smaller statistics and carry important information about the neutrino energy spectrum. 
\begin{figure}
\centering
    \includegraphics[width=0.47\textwidth]{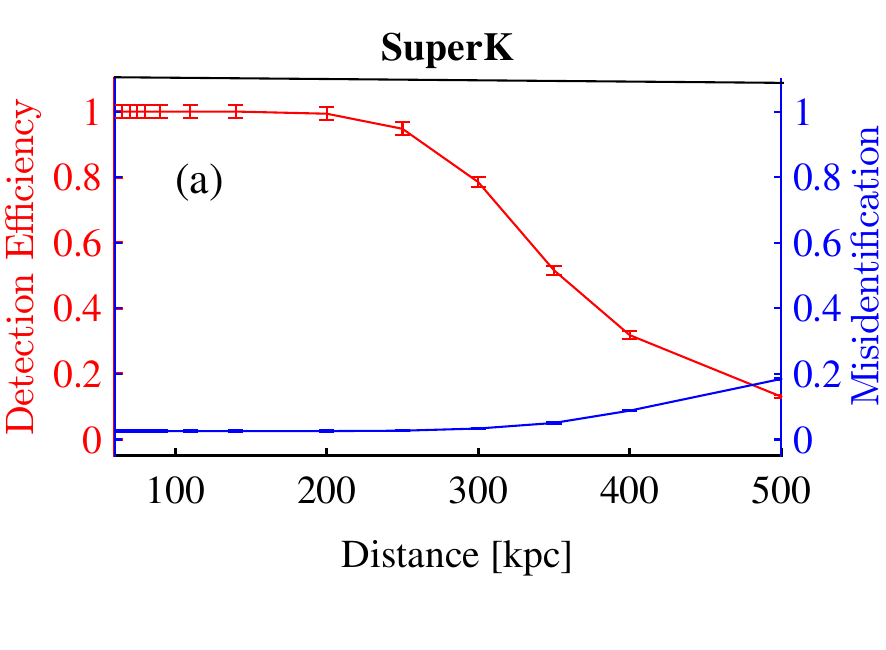}
    \includegraphics[width=0.47\textwidth]{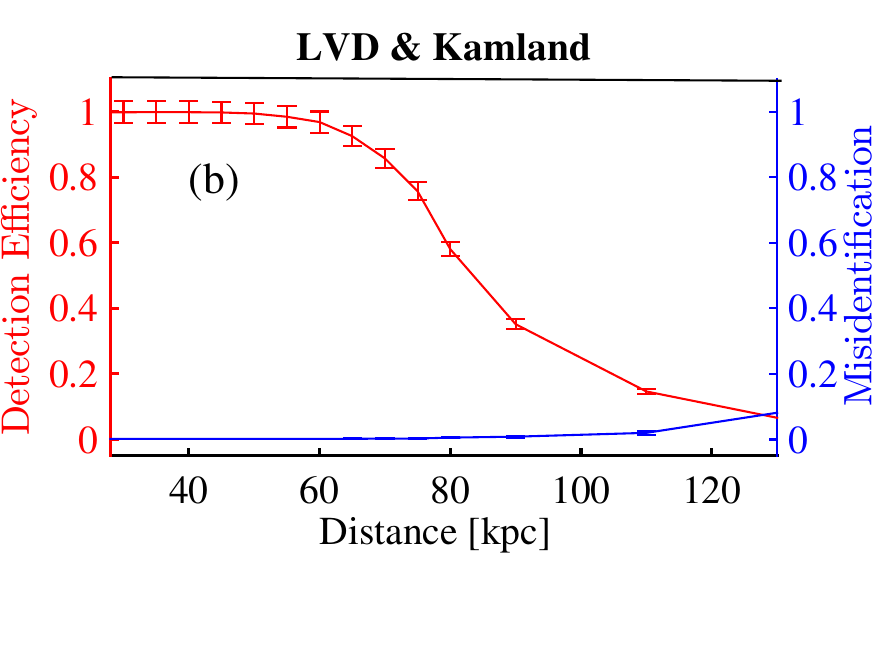}
    \caption{Detection efficiency and misidentification probability of neutrino detectors versus the CCSN distance. The panel (a) refers to Super-K while panel (b) show the results for LVD and Kamland working as a network. Adapted from C.~Casentini, G. ~Pagliaroli, C.~Vigorito and V.~Fafone, \textit{Pinpointing astrophysical bursts of low-energy neutrinos embedded into the noise}, JCAP \textbf{8} 10 (2018)\cite{Casentini:2018bdf}.\textcopyright~Casentini et al. Reproduced with permission.}
    \label{fig:efficiency_nu}
\end{figure}

Super-K, a $50$-kton water Cherenkov detector in Japan, could observe some ${\sim} 8000$ events for a core collapse at the center of the Milky Way, ${\sim} 8.5$~kpc away. The LVD 
and Borexino scintillation detectors at Gran Sasso in Italy, and KamLAND
in Japan, would observe hundreds of interactions from the same event. The IceCube detector at the South Pole, although nominally a multi-GeV neutrino detector, would observe a coincident increase in count rate in its phototubes due to a diffuse burst of Cherenkov photons in the ice, and has sensitivity to a galactic supernova; for a review see Ref. \cite{Scholberg:2012id} and references therein.

The duty factor of neutrino detectors in observing mode is typically $90\%$ or better and their horizon extends now up to the Large Magellanic Cloud with a very low misidentification probability, i.e., the probability that an events cluster due to background survives selection criteria as a signal \cite{Casentini:2018bdf}.    
Fig.~\ref{fig:efficiency_nu} reports the detection efficiencies of the Super-K (left panel) and the Kamland-LVD network(right panel). Super-K can reach an horizon of ${\sim} 200$ kpc with a detection efficiency of $\simeq 100\%$ and a misidentification probability of ${\sim} 3\%$. The Kamland and LVD detectors working together can reach the Large Magellanic Cloud with $100\%$ of detection efficiency and $0.2\%$ of misidentification probability.

The Super-K, LVD, IceCube, and Borexino detectors are also operating as a part of the SNEWS (SuperNova Early Warning System) network \cite{Antonioli:2004zb}, whose goal is to provide prompt alerts to astronomers in the case of a coincident supernova neutrino burst. Recently, the possibility of providing pre-SN alerts has also been discussed based on the expectations for the pre-CCSN neutrino signals. In the case of an extremely nearby CCSN ($D \sim 600$ pc), the Kamland detector can send an alert to prepare other detectors for observing upcoming signals \cite{Asakura:2015bga}.

Next generation neutrino detectors will increase the expected statistics of the supernova signal by scaling up the fiducial mass and increasing the horizon for the detection. For example, Hyper-Kamiokande (HK) \cite{Scholberg:2012id} could observe tens of events for a CCSN in Andromeda, which is ${\sim} 750$~kpc away. The Jiangmen Underground Neutrino Observatory (JUNO) \cite{An:2015jdp}, a $20$~kt underground liquid scintillator detector, will collect ${\sim} 2000$ all-flavor neutrino-proton elastic scattering events from a typical CCSN at a distance of 10 kpc.
The Deep Underground Neutrino Experiment (DUNE), $40$~kt LArTPC detector to be constructed underground in South Dakota, is expected to detect ${\sim} 3000$ events due to $\nu_e$ interaction from a $10$~kpc supernova, providing a $\nu_e$ sensitivity that complements the $\bar{\nu}_e$ interaction channel, dominant for most of other detectors\cite{Ankowski:2016lab}.

\subsubsection{Combined Searches}
\label{sec:mma.combined}
The multi-messenger nature of CCSNe motivates combined searches, i.e., searches performed by considering the information from different kinds of detectors and signals. In the simplest case, a CCSN detection (a $5 \sigma$ evidence) with a type of signal, e.g., an EM signal, is exploited for a correlated search for another type of signal expected from the same source, e.g., GWs. In the case of GW searches correlated with EM counterpart (optically targeted search), the EM detection imposes the sky location of the source, the source distance, and a broad time window for the arrival time of the GW signal (generally within a few days) \cite{Abbott:2019pxc}.
\begin{figure}
    \centering
    \includegraphics[width=\textwidth]{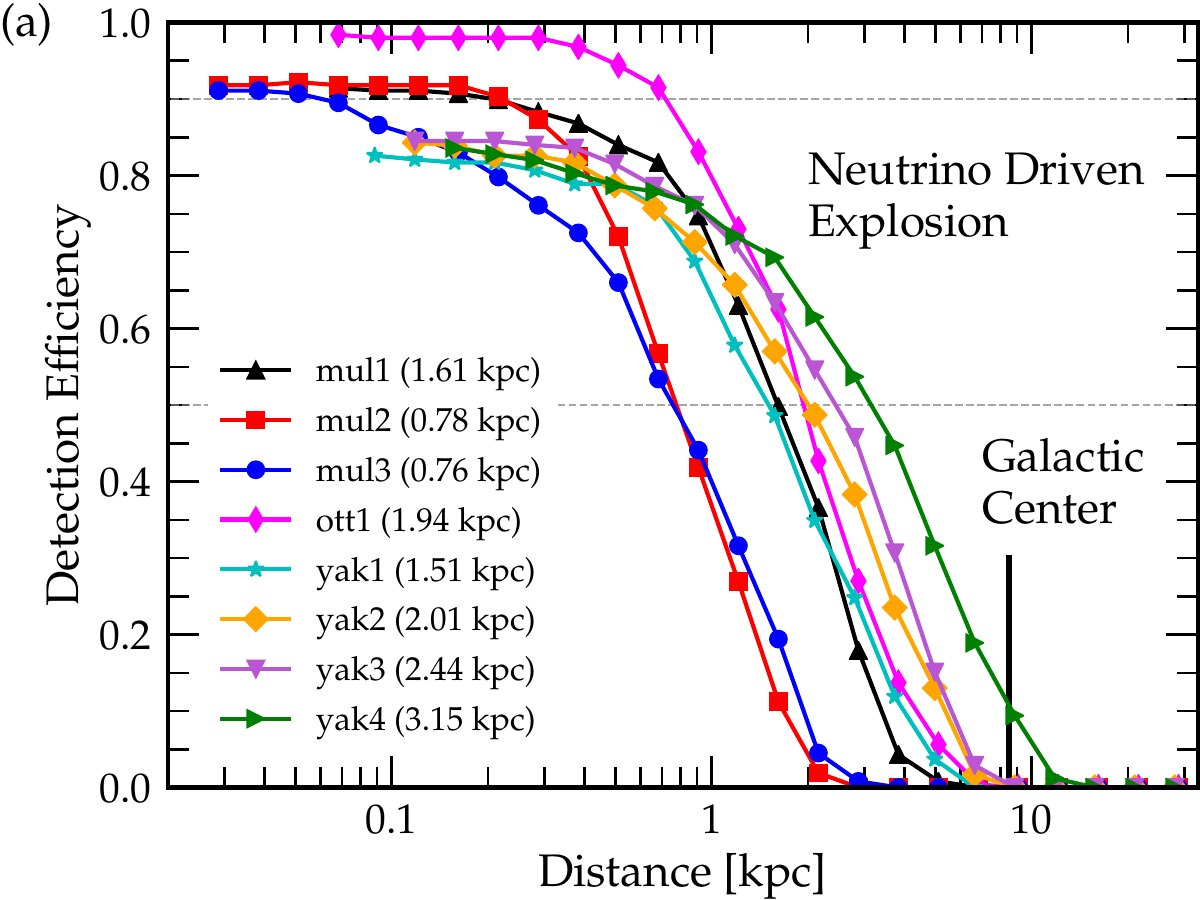}
    \caption{Detection efficiency as a function of the CCSN distance for a source located at the position and time of SN 2017eaw, i.e. with the perfect knowledge of the sky location and an on-source temporal window of about 1 day before the EM detection. Reproduced from B.P.~Abbott, and others, \textit{Optically targeted search for gravitational waves emitted by core-collapse supernovae during the first and second observing runs of advanced LIGO and advanced Virgo}, PRD \textbf{101} 8 (2020) \cite{Abbott:2019pxc}. \textcopyright~Abbott et al. Reproduced with permission.}
    \label{fig:efficiency}
\end{figure}
The detection efficiency can be estimated by injecting simulated signals in the CCSN optical position and looking for these signals by using the on-source window associated to the optical transient. In Fig.~\ref{fig:efficiency} as a leading example, the adopted source position is that of SN 2017eaw \cite{Abbott:2019pxc} and the observation window is about $1$ day. The injected waveforms are the results of different simulations, which considered the case of neutrino-driven explosions of non- or slowly rotating progenitor stars. In this scenario, GWs are emitted in the initial stage after the bounce in the frequency range 100-300~Hz due to prompt convection while, at later times, the frequency increases up to 2~kHz. Despite the fact that additional information provided by the optical counterpart increase the GW SNR, the detection horizon is less than $5$~kpc.  

The GW SNR enhancement can be stronger when a better suppression of background spikes is implemented by limiting the search to a small temporal window. This is in principle possible by exploiting the neutrino signals, which is characterized by a timescale much smaller than that of the EM signals. Indeed, the neutrino luminosity rises just after the core bounce and has been demonstrated that, by analyzing the $\bar{\nu_e}$ signal (IBD channel) from a galactic CCSN with the current generation of neutrino detectors, one can identify CCSN core ''bounce'' time within a window of $\sim 10$~ms or less \cite{Pagliaroli:2009qy}. This temporal window can be used to optimize a “triggered GW search”, i.e. to open an ad hoc temporal window (customized to fit the GW emission process around the time of the bounce) to look for GW bursts. The factor of improvement that can be achieved has been investigated in Ref.~\cite{Nakamura:2016kkl} for a CCSN located at the Galactic Center that emits the signals reported in Fig.~\ref{fig:multimessenger}. The CCSN coordinated observation with Super-K enables to restrict the time window to $[0,60]$ ms from the estimated time of the bounce. This increases the reconstructed SNR from ${\sim} 3.5$ to ${\sim} 7.5$, expanding the detection horizon in comparison to the one reported in Fig.~\ref{fig:efficiency} by approximately a factor of two.

In addition, using neutrino-electron forward scattering events, a Galactic core-collapse event can be pinpointed to within an error circle of some $5^{\circ}$ with the current Super-K detector \cite{Scholberg:2012id}. This uncertainty can be reduced by tagging the inverse beta decay events to $\sim 3^\circ$ with a tagging efficiency of 95\%, and, for large-scale detector as HyperK, this accuracy can be as good as $\sim 0.6^\circ$ \cite{Scholberg:2012id}. The DUNE detector is also expected to provide a comparable sensitivity \cite{Ankowski:2016lab}. The pointing capability critically impacts the prospects for the detection of the SBO \cite{Nakamura:2016kkl}.

For extra-galactic CCSNe, both the GW signal amplitude and the neutrino statistics are low. In these cases, the identification of the astrophysical bursts embedded into the detectors noise could be challenging when using GWs or neutrinos alone. However, by exploiting the temporal correlation among the neutrino and GW emissions, it is possible to perform a joint search of distant supernovae. 
A joint GW-neutrino search would enable improvements to searches by lowering the detection thresholds, increasing the distance ranges, and increasing the significance of candidate detections \cite{Leonor:2010yp}. 
Furthermore, neutrino experiments can benefit from a contemporary GW detection by relaxing the detection criteria. For example, Super-K's search of “distant” CCSNe requires two neutrino events (with energy threshold 17 MeV) within 20 s in order to have the accidental rate less than one per year. With these requirements the probability of detecting a supernova in Andromeda is approximately $8\%$. By requiring the coincidence of a single neutrino event with a gravitational wave signal the accidental rate could still be less than one per year, but the probability of detecting a core-collapse event in Andromeda would increase to $35\%$.

As last remark, it is important to stress that correlated signatures of GWs and neutrino emission can go beyond the pure temporal coincidence, as reported by numerical simulations \cite{Kuroda:2017trn}. In the case of successful explosions, it has been demonstrated that both the neutrino counts and GW evolution are correlated with the core compactness, the PNS mass at $1$~sec.~post-bounce, and the explosion energy\cite{GalloRosso:2017hbp}. However, for failed supernova events, both the neutrino and GW observable are correlated with the core compactness and the remnant BH mass \cite{Warren:2019lgb}. Deep studies of these astrophysical correlations can be essential to maximize the science return from such a rare event. 

\section{Conclusion and Prospects}
\label{sec:conclusions}

In this chapter, we summarized our current understanding of the gravitational wave (GW) emission from core-collapse supernovae (CCSNe). As discussed above, the GWs are predominantly produced by the oscillations of the protoneutron star (PNS). In the most common case of slowly rotating CCSNe, the multi-dimensional hydrodynamic flows that develop in the post-shock region perturb the PNS and excite its oscillations. The resulting signal is marginally observable with current detectors for a source within our galaxy, but future third-generation detectors will enable more detailed observation. In rare rapidly rotating progenitors, the PNS is born with a centrifugal deformation, which excites its "ring-down" oscillations that last for $\sim 10-20\,\mathrm{ms}$. Some rapidly rotating models may develop non-axisymmetric instabilities leading to long-lasting emission of GWs that could be detectable up to megaparsec distances. 

Despite the significant progress achieved so far, there is much more to do in the near future. Our ability to detect and interpret the GW signal from the next galactic supernova rests not only on the availability of ground-based laser interferometers, but also on the development of sophisticated data analysis techniques and reliable theoretical predictions. Progress on all of these crucial components is essential to turn a once in a lifetime event into a ground-breaking discovery.

On the data analysis side, techniques for the search of unmodelled signals could be complemented with methods that incorporate theoretical predictions based on PNS perturbation theory. Coincidence analysis of GW and neutrino burst signals should also be developed as a way not only to increase the sensitivity of the searches and decrease their false alarm rate, but also to test specific scenarios, such as the presence of correlated modulations of both signals due to SASI, that are predicted by state-of-the-art simulations.

Theoretical models need to be improved on several fronts. Most obviously, GW signals should be computed using simulations that consistently treat the gravity sector using general relativity, instead of using ad hoc prescription, such as the use of effective GR potentials in Newtonian calculations, as is done even in some of the most sophisticated simulations to date. Perhaps more importantly, even the longest simulations stop shortly after the onset of explosions, typically well before the end of the GW signal. Long-term simulations are needed to quantify the GW signal due to the continued accretion and the PNS inner convection. The impact of neutrino oscillations remains largely unexplored. While the direct impact of neutrino oscillations on the GW signal is expected to be small, neutrino oscillations need to be taken into account in the joint analysis of GW and neutrino data, and might even impact the dynamics of the explosion, thus affecting the overall signal. Finally, simulations of larger set of progenitors, including the ones with varying degree of rotation, will reveal the dependence on initial parameters.  

CCSNe occur once or twice per century in a galaxy, but the next galactic supernova has already happened. The EM, neutrino, and GW bursts from the event are already on their way toward the Earth, poised for the first coincident observation of neutrinos, photons, and gravitons by the international network of observatories. Their detection will bring a once in a lifetime scientific revolution in our understanding of the end of life of massive stars. With modern neutrino detectors, we will be able to directly observe and measure, among others, the core bounce and the associated breakout burst, the time variability of accretion, and the binding energy of the newborn NS. In the same way in which helioseismology revolutionized our understanding of the sun, GWs from the PNS will allow us to learn about the interior structures of NSs and constrain the turbulent nature of the accretion flows in the core of exploding stars. Finally, EM observations will allow us to learn about the pre-SN structure of massive stars and the degree of asymmetry in the explosion. In conclusion, CCSNe are the ultimate multimessenger source.

\section{Cross-References}
\begin{itemize}
    \item Ch. 1, Introduction to gravitational wave astronomy
    \item Ch. 2, Ground-based Laser Interferometers
    \item Ch. 7, 3rd generation of GW detectors
    \item Ch. 10, Isolated neutron stars
    \item Ch. 18, Electromagnetic counterparts of gravitational waves in the Hz-kHz range
    \item Ch. 19, Multi-messenger astronomy
\end{itemize}

\section{Acknowledgements}
We thank Adam Burrows and Kei Kotake for helpful comments. The work of G.P. was partially supported by the research grant number 2017W4HA7S ''NAT-NET: Neutrino and Astroparticle Theory Network'' under the program PRIN 2017 funded by the Italian Ministero dell'Istruzione, dell'Universita' e della Ricerca (MIUR).

\bibliographystyle{spphys}

\end{document}